%% file: main.tex
\documentclass[sigconf]{acmart}

\newif\ifcomments
\commentsfalse

\input{packages}
\input{macros}

\input{data_macros}

\copyrightyear{2025}

\acmYear{2025}
\acmVolume{YYYY}
\acmNumber{X}
\acmDOI{XXXXXXX.XXXXXXX}
\acmISBN{}

\begin{document}

\title{User Profiles: The Achilles' Heel of Web Browsers}


\author{Dolière Francis Somé}
\email{doliere.some@cispa.de}
\affiliation{%
  \institution{CISPA Helmholtz Center for Information Security}
  \city{Saarbr\"ucken}
  \state{Saarland}
  \country{Germany}
}

\author{Moaz Airan}
\email{mouazf.97@gmail.com}
\affiliation{%
  \institution{Saarland University}
  \city{Saarbr\"ucken}
  \state{Saarland}
  \country{Germany}
}

\author{Zakir Durumeric}
\email{zakir@cs.stanford.edu}
\affiliation{%
  \institution{Stanford University}
  \city{Stanford}
  \state{California}
  \country{USA}
}

\author{Cristian-Alexandru Staicu}
\email{staicu@cispa.de}
\affiliation{%
  \institution{CISPA Helmholtz Center for Information Security}
  \city{Saarbr\"ucken}
  \state{Saarland}
   \country{Germany}
}

\newcommand{\multilinebox}[1]{\pbox{\linewidth}{\vspace{.5\b‌aselineskip}#1\vspac‌e{.5\baselineskip}}}

\begin{abstract}
\input{abstract}
\end{abstract}

\keywords{Browser profiles, file system attackers, file system access API}

\maketitle

\input{introduction}
\input{background}
\input{fsa_api}

\input{threatmodel}

\input{fs_vs_browser_profiles}

\input{fsa_vs_browser_profiles}
\input{ustudy}

\input{countermeasures}
\input{disclosure}
\input{discussions}

\input{relatedwork}
\input{conclusion}

\bibliographystyle{ACM-Reference-Format}
\bibliography{main}

\appendix
\onecolumn
\input{appendix}

\end{document}

%% file: packages.tex
\usepackage{enumitem}
\setlist{nolistsep}
\usepackage{graphicx} 
\usepackage{hyperref}
\usepackage{wasysym}

\usepackage{tikz}
\usepackage{tablefootnote}
\usepackage{longtable}
\usepackage[linesnumbered,ruled,vlined]{algorithm2e}

\usepackage{textcomp}
\usepackage{booktabs}

\usepackage{makecell}

\usepackage{graphicx}
\usepackage{listings}
\usepackage{color}
\usepackage{caption}

\usepackage{footnote}

\usepackage{array}
\usepackage{mathpartir}
\usepackage{amsthm}
\usepackage{stmaryrd}
\usepackage{diagbox}
\usepackage{multirow}
\usepackage{subcaption}
\usepackage{csquotes}
\usepackage{colortbl}
\usepackage{xurl}
\usepackage{algpseudocode}
\usepackage[linesnumbered,ruled,vlined]{algorithm2e}
\usepackage{pifont}
\usepackage{fontawesome}
\usepackage{pdfpages}
\usepackage{listings}

%% file: macros.tex
\ifcomments
    \newcommand{\crs}[1]{[{\color{blue}Cris: #1}]}
    \newcommand{\fra}[1]{[{\color{red}Francis: #1}]}
\else
    \newcommand{\crs}[1]{}
    \newcommand{\fra}[1]{}
\fi

\def\home{\emph{\$HOME/}}
\def\homedir{\emph{$~$/}}
\def\username{\code{Username}}
\newcommand{\sx}[1]{(\S\ref{#1})}

\definecolor{lightgray}{rgb}{0.95, 0.95, 0.95}
\definecolor{darkgray}{rgb}{0.4, 0.4, 0.4}
\definecolor{purple}{rgb}{0.65, 0.12, 0.82}
\definecolor{editorGray}{rgb}{0.95, 0.95, 0.95}
\definecolor{editorOcher}{rgb}{1, 0.5, 0} 
\definecolor{editorGreen}{rgb}{0, 0.5, 0} 

\def\code#1{\texttt {\small{#1}}}

\def\REMARK#1#2{        
  \ifremarks
  \begin{center}
  \noindent\fbox{
  \begin{minipage}[b]{0.4\textwidth}
  \textbf{[#1]: #2}
  \end{minipage}}
  \end{center}
  \fi}

\def\todo#1{\REMARK{TODO}{\color{red}{#1}}}
\def\done#1{\REMARK{DONE}{\color{green}{#1}}}
\def\done#1{}

\newcommand{\greenc}{}%
\DeclareRobustCommand{\greenc}{%
  \tikz\fill[scale=0.4, color=green]
  (0,.35) -- (.25,0) -- (1,.7) -- (.25,.15) -- cycle;%
}

\def\greenc{\CIRCLE}
\def\redc{\Circle}
\def\grayc{\LEFTcircle}

\def\greencheck{\ding{52}}
\def\redcheck{\ding{55}}

\definecolor{Gray}{gray}{0.9}
\definecolor{LightCyan}{rgb}{0.88,1,1}
\definecolor{DarkGray}{rgb}{0.9,1,1}
\definecolor{Silver}{rgb}{192,192,192}
\definecolor{Maroon}{rgb}{128,0,0}
\definecolor{Teal}{rgb}{0,128,128}
\definecolor{Olive}{rgb}{128,128,0}
\definecolor{Navy}{rgb}{0,0,128}
\definecolor{Orange}{rgb}{255,165,0}
\definecolor{SeaGreen}{rgb}{46,139,87}
\definecolor{Wheat}{rgb}{245,222,179}
\definecolor{Ivory}{rgb}{255,255,240}
\definecolor{Azur}{rgb}{240,255,255}
\definecolor{RoyalBlue}{rgb}{65,105,225}
\definecolor{Salmon}{rgb}{250,128,114}

\lstdefinelanguage{CSS}{
  keywords={color,background-image:,margin,padding,font,weight,display,position,top,left,right,bottom,list,style,border,size,white,space,min,width, transition:, transform:, transition-property, transition-duration, transition-timing-function}, 
  sensitive=true,
  morecomment=[l]{//},
  morecomment=[s]{/*}{*/},
  morestring=[b]',
  morestring=[b]",
  alsoletter={:},
  alsodigit={-}
}

\lstdefinelanguage{CSP}{
  keywords={default-src, script-src, child-src, frame-src, script-src, style-src, report-uri, connect-src, img-src, object-src, frame-ancestors, manifest-src, plugin-types, form-action, sandbox, worker-src, font-src, media-src},
  morestring=[b]',
  alsoletter={:},
  alsodigit={-}
}

\lstdefinelanguage{HTTP}{
  keywords={Location, Content-Security-Policy, Content-Security-Policy-Report-Only},
  morestring=[b]',
  alsoletter={:},
  alsodigit={-}
}

\lstdefinelanguage{FP}{
  keywords={camera, geolocation, midi, document-domain, microphone, payment, sync-xhr},
  morestring=[b]',
  alsoletter={:},
  alsodigit={-}
}

\lstdefinelanguage{XFO}{
  keywords={SAMEORIGIN, DENY},
  morestring=[b]',
  alsoletter={:},
  alsodigit={-}
}

\lstdefinelanguage{RP}{
  keywords={no-referrer, same-origin, origin, strict-origin, strict-origin-when-cross-origin, origin-when-cross-origin, unsafe-url, no-referrer-when-downgrade},
  morestring=[b]',
  alsoletter={:},
  alsodigit={-}
}

\lstdefinelanguage{JavaScript}{
  keywords={break, case, catch, continue, debugger, default, delete, do, else, finally, for, function, if, in, instanceof, new, return, switch, this, throw, try, typeof, var, void, while, with},
  keywordstyle=\color{blue}\bfseries,
  ndkeywords={class, export, boolean, throw, implements, import, this, true, false},
  ndkeywordstyle=\color{darkgray}\bfseries,
  identifierstyle=\color{black},
  morecomment=[s]{/*}{*/},
  morecomment=[l]//,
  morestring=[b]",
  morestring=[b]'
}

\lstdefinelanguage{Python}{
  keywords={def, break, case, except, continue, global, del, do, else, elif, for, if, in, new, return, switch, this, throw, try, while, with},
  keywordstyle=\color{blue}\bfseries,
  ndkeywords={class, export, import, this, True, False},
  ndkeywordstyle=\color{darkgray}\bfseries,
  identifierstyle=\color{black},
  morecomment=[s]{/*}{*/},
  morecomment=[l]//,
  morestring=[b]",
  morestring=[b]'
}
\lstdefinelanguage{HTML5}{
  language=html,
  sensitive=true, 
  alsoletter={<>=-},  
  morecomment=[s]{<!-}{-->},
  tag=[s],
  otherkeywords={
  >,
  <!DOCTYPE,
  </html, <html, <head, <title, </title, <style, </style, <link, </head, <meta, />,
  </body, <body,
  </div, <div, </div>, 
  </p, <p, </p>,
  </script, <script,
  <canvas, /canvas>, <svg, <rect, <animateTransform, </rect>, </svg>, <video, <source, <iframe, </iframe>, </video>, <image, </image>
  },
  ndkeywords={
  =,
  charset=, src=, id=, width=, height=, style=, type=, rel=, href=,
  fill=, attributeName=, begin=, dur=, from=, to=, poster=, controls=, x=, y=, repeatCount=, xlink:href=,
  margin:, padding:, background-image:, border:, top:, left:, position:, width:, height:,
  transform:, -moz-transform:, -webkit-transform:,
  animation:, -webkit-animation:,
  transition:,  transition-duration:, transition-property:, transition-timing-function:,
  }
}

\newif\ifreport\reporttrue 
\reportfalse

\newif\ifpaper\papertrue 
\newif\ifremarks\remarksfalse
\remarkstrue 
\newif\ifdraft\draftfalse 

%% file: data_macros.tex
\def\chromiumfirefoxwebkitmarketshare{97.36\%}
\def\flatpak{\emph{flatpak}}

\def\participantsfinal{38}
\def\downloadzip{34}
\def\downloadfsa{4}

\def\uploadfsa{23}
\def\uploadtraditional{15}

\def\downloadzipuploadfsa{19 (50\%)}

%% file: abstract.tex
Web browsers provide the security foundation for our online experiences. Significant research has been done into the security of browsers themselves, but relatively little investigation has been done into how they interact with the operating system or the file system.
In this work, we provide the first systematic security study of browser profiles, the on-disk persistence layer of browsers, used for storing everything from users' authentication cookies and browser extensions to certificate trust decisions and device permissions. We show that, except for the Tor Browser, all modern browsers store sensitive data in home directories with little to no integrity or confidentiality controls. 
We show that security measures like password and cookie encryption can be easily bypassed. In addition, HTTPS can be sidestepped entirely by deploying malicious root certificates within users' browser profiles. The Public Key Infrastructure (PKI), the backbone of the secure Web—HTTPS—can be fully bypassed with the deployment of custom potentially malicious root certificates. 
More worryingly, we show how these powerful attacks can be fully mounted directly from web browsers themselves, through the File System Access API, a recent feature added by Chromium browsers that enables a website to directly manipulate a user's file system via JavaScript. In a series of case studies, we demonstrate how an attacker can install malicious browser extensions, inject additional root certificates, hijack HTTPS traffic, and enable websites to access hardware devices like the camera and GPS. 
Based on our findings, we argue that researchers and browser vendors need to develop and deploy more secure mechanisms for protecting users' browser data against file system attackers.



%% file: introduction.tex
\section{Introduction}
\label{sec:introduction}

%

To mediate users' interactions with the web, browsers have deployed dozens of security mechanisms. This includes sandboxing untrusted Javascript code~\cite{SOP}, hardening browser media renderers to isolate websites from one another~\cite{CORS}, and fine-grained permissions mechanisms for accessing device hardware~\cite{CSP}. 
The academic community published hundreds of papers studying the limits~\cite{LuoLHN19, Nomoto0SAM23, SnyderKELH23} or testing~\cite{BernardoVVCSAM24,ShouKSB21,ZhouZWGLLP022,BrownSE20} defenses for web browsers, leading to significant improvements.  
Despite the plethora of work analyzing how browsers {internally} operate and how they can defend against attacks targetting visited websites during user's browsing sessions, there has been relatively little work investigating \textit{how resilient browsers are against attackers that have access to the file system}. Traditionally, such attackers were considered too strong and deemed out of scope in browsers' threat models. However, recent developments urge us to consider this threat model for modern browsers. Below, we list three essential supporting arguments. 

First, the advent of the File System Access (FSA) API in Chromium browsers allows web attackers to access the file system directly. Oz et al.~\cite{Oz-etal-23-USENIX} show that this API can be misused for ransomware attacks, but they argue that it cannot be abused for stealing browsers' data because of the blocklist mandated by the FSA standard~\cite{FSAAPI-Block}. However, as noted by the Chromium team~\cite{Chromium-issue-discussion,chromesecfaq}, this blocklist is not a silver bullet but a pragmatic defense against blatant misuse. In this work, we 
demonstrate that remote attackers can leverage the FSA API to compromise data stored by browsers on the disk.
Second, the landscape of unwanted software~\cite{Urban-etal-18-ESORICS} has changed significantly in recent years with the advent of prevalent supply chain attacks~\cite{OhmPS020}. 
Traditionally, browser vendors often consider defenses against malware present on the machine out of scope~\cite{chromesecfaq}, arguing that such attackers can mount more powerful attacks, such as altering the browser's binaries. However, there are several recent reports of open-source packages stealing information persisted by browsers on the disk: passwords~\cite{steal-passwords}, session cookies~\cite{steal-cookies}, entire browser profiles~\cite{steal-profiles}, extensions' data~\cite{steal-ext-data}. These malicious packages have complete power over the victim's machine, but they specifically target these browser artifacts, allowing them to carry out web attacks~\cite{infostealingmalware,cookietheftmalware}. 
Third, we already see initial attempts by the vendors to deploy defenses against file system attackers. For example, as discussed in Section~\ref{sec:background}, Chromium-based browsers and Firefox encrypt user passwords before storing them on the disk. Picazo-Sanchez et al.~\cite{Sanchez-etal-20-CANS} also discuss how a simple HMAC mechanism in Chromium browsers is implemented to protect preference files and extensions' integrity. Our paper shows that these mechanisms are insufficient against file system attackers. Notably, in concurrent and independent work, browser vendors are exploring more principled solutions for protecting their persistence layer, using device-bound session credentials (DBSC)~\cite{DBSC-Chromium} that aims to prevent cookie theft or application-bound encryption on Windows~\cite{App-bound-enc-Windows}. While these proposals show that browser vendors are taking file system attackers seriously, they are in an early phase and, thus, still to be widely adopted by all vendors and operating systems. 


%
%
%
Considering the vendors' interest and a high potential for securing and compromising browser profiles, 
we systematically analyze the data stored by modern browsers on the disk and ways of compromising it.
We answer the following research questions:

\noindent
{\bf RQ1: What data do browsers store in user-accessible file system locations, and what security mechanisms are implemented to protect it against external manipulation? (Section~\ref{sec:background})}
Web browsers store user-specific data and configuration files in \textit{browser profile folders}.  When a user first launches a browser, it creates and stores the user profile data in well-known locations depending on the operating system, browser, and configurations~\cite{ChromeProfileFolders, FirefoxProfileFolders}. 
A user's profile consists of an on-disk collection of databases, configuration, and preferences files that the browser writes to preserve the state of the browser. Depending on the browser, this profile contains data like installed extensions, stored usernames and passwords, cookies, bookmarks, custom root certificate authorities, proxy preferences, and permissions for accessing physical devices such as cameras or microphones. A subset of this data, including cookies and passwords, are encrypted on disk. Another subset, including browser extensions, has integrity checks. Other configurations, like device access and trusted certificate authorities, have no protections.
Unlike many application configuration files, we emphasize that a browser's configuration can fundamentally alter its security-relevant behavior. 
We find that any application that can write to a user's home directory (or equivalent) can easily install code into a browser, proxy traffic to attacker websites, and change device permissions simply by changing on-disk configuration files. 

\noindent
{\bf RQ2: What attacks can be mounted against browsers, users, and websites by tampering with profile folders? (Sections~\ref{sec:threatmodel} and \ref{sec:fs_vs_browser_profiles})}
We show that several attacks can be mounted by a file system adversary against browsers, assuming various access levels to a user's profile directory, i.e., read, write, or execute. We find that most security mechanisms, like encryption or integrity checks, can be bypassed by attackers with a write access to the profile directory. In addition, we demonstrated that it is possible to mount these attacks remotely through the FSA API, with JavaScript in websites. Combining read and write access to browser profiles, we showcase serious security and privacy implications with a series of end-to-end attacks, ranging from (i)~a complete hijack of Firefox browsers, (ii)~the installation of malicious extensions in Chromium browsers, (iii)~bypassing encryption on cookies and saved login passwords, (iv)~installing malicious root certificate authorities, (v)~redirecting user traffic to a man-in-the-middle (MiTM) proxy, and (vi)~silently spying on the user by enabling their camera, microphone, and GPS. 

\noindent
{\bf RQ3: How can we practically improve browsers' security against file system attackers? (Section~\ref{sec:countermeasures})}
We argue that browsers can adopt several practical protections to protect user profiles, particularly against attacks via the FSA API. Several major vendors indicated methods for safeguarding the profiles in our disclosure process. In contrast, others noted that they believed our presented attacks were out-of-scope and argued that the file system's security is the operating system's mandate, not the browser's. 
We strongly believe that browsers should ensure that persisted profiles on the disk are machine-bound, opaque, and immutable to other applications.
To this end, we advocate for implementing strong encryption schemes on browser profiles, which we demonstrate are practical. In addition, we suggest that operating systems provide a mechanism for checking signed apps to manage app data securely. 

%

Overall, we make the following contributions: 
\begin{itemize}
    \item We present the first systematic risk analysis of browser profiles, their data, and their susceptibility to compromises by a non-privileged file system attacker. We show that attacks can be remotely mounted via the JavaScript FSA API. We present end-to-end attacks, that can even be mounted by weak web adversaries. Doing so, we demonstrate that browser profiles compromise is a realistic threat. 
    \item We extensively compare modern browsers' susceptibility to browser profile compromise and study their protection mechanisms. Most data is stored in clear and is readily accessible to attackers. Moreover, most security mechanisms like confidentiality or integrity checks are bypassable.
    \item We discuss how browsers can practically prevent profile compromise attacks. We advocate for opaque, immutable, and immovable profiles. In other words, each persisted profile should be machine-bound, and there should be no way to read or modify only parts of the profile from outside the browser. We demonstrate and evaluate the feasibility of an encryption schema on top of browser profiles.
\end{itemize}
We refer the reader to the associated repository~\cite{PoC}, where we demonstrate, 
the attacks and defenses presented in this paper.

%% file: background.tex
\begin{table}[tp]
    \footnotesize{
    \centering
    \begin{tabular}{|p{2.5cm}|p{5.1cm}|}
        \hline
         \bfseries{Browser} & \bfseries{Profile folders default locations} \\ \hline 
            \multirow{3}{*}{Mozilla Firefox} & \code{\scriptsize .mozilla/firefox/} [L] \\ 
             &  \code{\scriptsize AppData\symbol{92}Roaming\symbol{92}Mozilla\symbol{92}Firefox\symbol{92}Profiles\symbol{92}} [W] \\ 
             &  \code{\scriptsize Libray/Application Support/Firefox/Profiles/} [M]\\ \hline
       \multirow{1}{*}{Webkit Safari} & \code{\scriptsize Libray/Safari/} [M]\\ \hline 
        \multirow{3}{*}{\parbox{2.5cm}{Google Chrome}} & \code{\scriptsize .config/google-chrome/} [L] \\ 
             &  \code{\scriptsize AppData\symbol{92}Local\symbol{92}Google\symbol{92}Chrome\symbol{92}User Data\symbol{92}} [W] \\ 
             &  \code{\scriptsize Libray/Application Support/Google/Chrome/} [M]\\ \hline
         \multirow{3}{*}{\parbox{2.5cm}{Microsoft Edge}} & \code{\scriptsize .config/microsoft-edge/} [L] \\ 
             &   \code{\scriptsize AppData\symbol{92}Local\symbol{92}Microsoft\symbol{92}Edge\symbol{92}User Data\symbol{92}} [W] \\ 
             &  \code{\scriptsize Libray/Application Support/Microsoft Edge/} [M]\\ \hline 
    \end{tabular}
    \caption{Default profile folder locations of major browsers on Linux (L), Windows (W), and macOS (M), relative to the user's home directory. The latter typically takes the form \code{/home/\username/}, \code{/Users/\username/}, \code{C:\textbackslash Users\textbackslash \username\textbackslash}.}
    \label{tab:browser_default_profile_folders}
    }
\end{table}

\begin{table}[tp]
    \footnotesize{
    \centering
    \begin{tabular}{|l|l|l|l|l|l|l|l|l|l|}
        \hline 
         & \multicolumn{2}{c|}{\bf Firefox} & \multicolumn{2}{c|}{\bf Chromium}  & \multicolumn{2}{c|}{\bf Safari} \\ \hline 
         & \textit{C} & \textit{I} & \textit{C} & \textit{I} & \textit{C} & \textit{I} \\ \hline
         \rowcolor{gray}\bfseries{Cookies} & \redcheck & \redcheck & \greencheck & \redcheck  & & \\ \hline 
        \rowcolor{lightgray}\bfseries{User Credentials (Passwords)} & \greencheck & \redcheck  & \greencheck & \redcheck  & \greencheck & \greencheck \\ \hline 
         \rowcolor{gray}\bfseries{Browser Extensions} & \redcheck & \greencheck & \redcheck & \greencheck  & & \\ \hline 
        \rowcolor{lightgray}\bfseries{Device Permissions} & \redcheck & \redcheck & \redcheck & \redcheck  & \redcheck & \redcheck \\ \hline 
         \rowcolor{gray}\bfseries{History/Bookmarks/Downloads/} & \redcheck & \redcheck & \redcheck & \redcheck & \redcheck & \redcheck  \\ \hline 
        \rowcolor{lightgray}\bfseries{User/Browser Preferences} & \redcheck & \redcheck & \redcheck & \greencheck & & \\ \hline 
         \rowcolor{gray}\bfseries{Root Certificates} & \redcheck & \redcheck  & \redcheck & \redcheck & & \\ \hline 
    \end{tabular}
    \caption{Excerpt of sensitive content stored in profiles and the presence (\greencheck) or absence (\redcheck) of security mechanisms deployed to protect profiles' confidentiality (\textit{C}) or integrity (\textit{I}).}
    \label{tab:excerpt_browser_profiles_content_and_security}
    }
\end{table}

\section{Browser Profiles}
\label{sec:background}
\label{sec:browser_profiles}
\label{sec:browserprofiles}
\label{sec:study_user_browser_profile}
Browsers read and write files that define or customize their runtime behavior and state, website data, and user-sensitive information. This information is organized in profile folders and stored on the disk. 
In this study, we focus on the three prominent desktop open-source browser families, i.e., Chromium-based or Chromium (e.g., Chrome, Edge, Opera, Vivaldi, Chromium, Brave), Gecko-based or Firefox browsers (Firefox, Waterfox, LibreWolf, Palemoon)\footnote{Tor is a Firefox-based browser, but it follows unique security and privacy practices that dictate that minimal to no data is stored in the user profile folder} and WebKit browsers (Safari and Orion). 
Cumulatively, these three browsers account for approximately~\chromiumfirefoxwebkitmarketshare\ of the desktop browsers' global market share at the time of writing (i.e., 03/2025)~\cite{DesktopBrowsersMarketShare}. The Chromium and Firefox browsers are cross-platform (work on major operating systems including Linux, Windows, and Mac OS), while Webkit browsers are mostly found on Mac OS. For clarity, we show code snippets and examples for a Linux system, specifically Ubuntu. 
\\
{\bf Default folders locations.}
On all operating systems, the user's home directory (e.g., \code{/home/\username/} on Linux, \code{/Users/\username/} on MacOS, or \code{C:\textbackslash Users\textbackslash \username\textbackslash} on Windows) is generally the containing root folder under which browsers create and store their data. Table~\ref{tab:browser_default_profile_folders} lists common default locations where major browsers store their profile folders. For simplicity, we often use \home\ or \homedir\ to refer to the user home directory. On Windows and macOS, the \code{AppData\symbol{92}} and \code{Library/} are respectively prominent subfolders (relative to the home folder) where one can find the browsing profile of most browsers. Linux has less uniformity (or more diversity), but \code{.config/, .mozilla/} are predominant locations of default Chromium and Firefox browsers profile folders. 
%
\\
{\bf Custom folders locations.} Browsers allow users to create additional browsing profiles and switch among them when the browser launches. This can, for instance, be done with command line options like \code{----user-data-dir} (Chromium) or \code{--profile} for Gecko browsers.
Browsers do not impose restrictions on the locations of the custom profile folders on the user's device. The \code{-P} command line flag of Firefox prompts a graphical user interface (GUI) or screen for managing profile folders. As for Safari, we are unaware of command line options that make it possible to create additional profile \emph{folders}, except from the browser's graphical settings.
\\
{\bf Profiles.}
Browsers launch on a profile \emph{folder} (default or custom), which typically contains a default and additional custom \emph{profiles}. One can view those as a way of isolating different browsing sessions, e.g., separate cookies corresponding to various sessions. At another level, profiles under the same folder share the same settings, like the list of installed web extensions, while different folders typically do not have common settings or preferences. For Gecko-browsers, by navigating the special \code{about:profiles} URI, one can view the current profiles and folders in use and further manage them. Chromium browser profile folders can be viewed at \code{chrome://version}. In the settings menu, Chromium browsers also offer the possibility to operate additional profiles\footnote{https://support.google.com/chrome/answer/2364824}. As for Safari, in addition to the default profile, starting from version 17, it is possible to create additional profiles\footnote{https://support.apple.com/en-us/105100}. Interestingly, it is the only browser that makes it possible to have multiple profiles in mobile browsers.
\\
{\bf Profiles folders structure.}
The path to the \emph{default} profile for the Chrome browser has the form \textit{\$HOME/.config/google-chrome/Default/}, and custom profiles follow a similar schema. For Firefox, it has the form \textit{\$HOME/.mozilla/firefox/$RANDOM$.default-release}, where \textit{RANDOM} is a random string, e.g., 8sypm4uk (See Figure~\ref{fig:fsa_file_browsing}), while custom profiles also follow a similar structure. As for Safari, managing these profiles requires manual user interactions, and the browser chooses the location where they are stored.
\\
{\bf Profiles folders' content and security mechanisms.}
In general, profile folders contain information such as the saved user credentials (usernames and passwords), cookies, bookmarks, custom certificate authorities, HTTPS proxy preferences, permissions to access physical devices such as the camera, microphone, GPS, or the list of installed extensions/add-ons, etc. Due to page limits, we only discuss the content of the profile folders relevant to the security and privacy issues we later demonstrate in this work, providing details about the content and, whenever applicable, the security mechanisms implemented by browsers to protect the content in the folder. Table~\ref{tab:excerpt_browser_profiles_content_and_security} shows an excerpt of typical content in profile folders and the security mechanisms browsers implement to protect them. We observe that the amount and format of data in profile folders are relatively similar among Gecko and Chromium desktop browsers and focus on these two browser families. We refer the reader to Section~\ref{sec:discussion} for a discussion about the Safari browser. This browser was unaffected by most of the attacks considered in this work.  

\subsection{Cookies}
\label{sec:browser_profiles_cookies}
%
Cookies add state to the otherwise stateless HTTP protocol and are leveraged for user session management, as well as for user tracking and the delivery of targeted advertisements. Access to cookies allows attackers to perform attacks like session hijacking, session fixation, or cross-site request forgery (CSRF) attacks. To tackle these issues, cookies have been enhanced over time with various security tags, flags, and prefixes like \emph{HttpOnly}, which hides the cookies to (potentially malicious) JavaScript programs, or the \emph{Secure} flag, which instructs the browser to send cookies only over HTTPS connections to prevent their interception by a network attacker. Recent features include the \emph{SameSite} and other prefixes (\emph{\_\_Host-} and \emph{\_\_Secure-}) that aim at mititating session fixation~\cite{CookiesMDN}.
\\
{\bf Cookies in browser profiles.}
In both browser families, cookies are stored in SQLite databases, i.e., \code{cookies.sqlite} for Firefox and \code{Cookies} for Chromium. The different tables, i.e., \code{moz\_cookies} (Firefox) and \code{cookies} (Chromium) and their columns store and describe the cookies set by websites along their attributes, including the domain, name, value, and flags and prefixes like \emph{HttpOnly}, \emph{Secure}.
\\
{\bf Security mechanisms.}
Cookies are stored \emph{in clear} in Firefox browsers, making them trivially accessible to attackers with a read capability. In Chromium browsers, the cookie \emph{values} are stored \emph{encrypted}. Section~\ref{sec:attacks_cookies} demonstrates a bypass of this encryption. 

\subsection{User Credentials}
\label{sec:browser_profiles_credentials_and_passwords}
%
Until the Web successfully turns to passwordless solutions in the upcoming years, most web applications rely on passwords as their services' primary access control mechanisms. In exchange for a valid username/password (credentials) and potentially additional authentication factors, the browser receives and stores session cookies that can be used to protect parts of web applications. As a critical topic in web security, different metrics have been devised to ensure credentials' security: passwords must be kept secret, unique per site, and strong in terms of entropy and length to resist dictionary and brute force attacks. As a result, users have to remember dozens of random passwords. To aid with this, browsers offer users the possibility to save the passwords and auto-fill them on associated websites. Standalone password managers packaged as browser extensions or native apps are popular alternatives, but we focus on built-in browser password managers in this work.
\\
{\bf Passwords in browser profiles.}
In Chromium browsers, passwords are encrypted and stored in the \code{Login Data} SQLite database under the table \code{logins}. In Firefox browsers, user credentials are encrypted and stored in the \code{logins.json} JSON file. Additionally, Firefox stores in the SQLite \code{key4.db} database the primary key (which is further encrypted) that is first salted and used to decrypt the credentials with a 3DES encryption algorithm. By default, the key is empty unless the user sets one. In Section~\ref{sec:attacks_saved_credentials_and_passwords}, we demonstrate that an attacker can bypass the encryption of user credentials.

\subsection{Browser Extensions}
\label{sec:browser_profiles_extensions}
%
Browser extensions are third-party programs that extend browser functionality and enhance users' browsing experiences. In this work, we focus on the WebExtensions cross-browser extensions API supported by major browsers, including Chromium, Gecko, and WebKit~\cite{WebExtensionsMDN, WebExtensionsChrome, WebExtensionsSafari}. Among other capabilities, extensions can inject code in webpages with content scripts and web-accessible resources, intercept and manipulate their requests and responses with the \emph{webRequest} API, or redirect traffic to proxy servers with the \emph{proxy} API, etc. Given their access to elevated privileges and APIs in the browser, the process of publishing and installing extensions is guarded by drastic security mechanisms. First, extensions are signed and must come from trusted sources, like the CWS~\cite{CWS-Chrome} and AMO~\cite{AMO-Firefox}. During the installation process, users are prompted to review and confirm the permissions requested by the extensions to operate correctly at runtime. While downloading the extension from trusted sources, the browser performs integrity checks to detect manipulations during transit (e.g., by a network attacker). 
\\
{\bf Extensions in profile folders and security.}
Once installed, the extension sources (JavaScript, HTML, CSS, images, etc.) are saved in the profile folder under \code{extensions/} for Firefox and \code{Extensions/} for Chromium. In the case of Firefox, the extension's original ZIP file (\code{.xpi}) is stored. On Chromium, however, the unpacked extension resources are stored on disk. Additional metadata and security files are generated and stored along the extensions. For Firefox, the metadata is part of the ZIP bundle in the \code{META-INF/} subfolder. For Chromium, the metadata is stored under the unpacked extensions root folder under \code{\_metadata/}. Additional signatures related to installed extensions are also stored in the \code{Secure Preferences} file~\cite{Sanchez-etal-20-CANS}. Finally, each time the browser launches, the local extension sources go through integrity checks to ensure they have not been tampered with. If this is the case, the extension may be offloaded and reinstalled. 
In Section~\ref{sec:attacks_extensions}, we bypass the integrity check of installed extensions and stealthily add new malicious ones. 

\subsection{Users Permissions}
\label{sec:browser_profiles_permissions}
%
Websites' ability to access device resources such as storage, camera, microphone, GPS, and web push notifications is guarded by explicit permission from the user, all mediated by the browser. Indeed, users trust browsers and usually allow them to access these devices and further delegate access to websites upon request. When a website asks permission to access a feature, a pop-up window asks for the user's consent. Once the user grants permission, the website can access the associated device, and the browser usually remembers this choice for future requests. 
\\
{\bf Devices access permissions in profile folders.}
In Firefox, all permissions are stored in an SQLite database (\texttt{permissions.sqlite}). Each entry consists of a domain, name of permission, a number representing whether the permission was allowed or blocked, and other time info. In Chromium, all permissions are stored in the \code{Preferences} JSON file. 
In Safari browsers, permissions are stored in the \emph{UserNotificationPermissions.plist} and \emph{UserMediaPermissions.plist} files. Decoding these encoded Safari browser files with the \code{plutil} command yields an XML file with names of websites and permissions they requested.
Changes made to these permissions databases will be blindly trusted and applied by the browser when it runs. In Safari, these permissions affect all the profiles, while in other browsers, permissions are granted on a per-profile basis. Section~\ref{sec:attacks_permissions} illustrates attacks against permissions in browser profiles.

\subsection{Custom Root Certificates}
\label{sec:browser_profiles_mitms}

The Public Key Infrastructure (PKI) is the backbone that ensures secure communications on the Internet, providing confidentiality, integrity, and authentication of parties (clients and servers). With the HTTPS protocol, browsers come bundled with a limited set of trusted root certificates, which they interrogate to check the validity of certificates presented by web servers the user intends to connect to. Becoming a trusted root certificate is a privilege that can be (ab)used to generate and successfully pass the verification of certificates for random websites and potentially mount man-in-the-middle (MiTM) attacks. To mount a MiTM attack where all requests and responses are redirected to an HTTPS proxy, one has to be able to (i) make the browser trust a custom root certificate authority (CA) and (ii) additionally instruct the browser to redirect traffic to an authoritative MiTM proxy server operating the malicious CA. 
\\
{\bf Custom root certificates in browsers.}
Major browsers offer the possibility to specify custom root certificates. In Firefox browsers, the \code{cert9.db} and \code{key4.db} files that contain custom CAs are directly located under the browsing profile. To specify the proxy server to which HTTPS traffic is redirected, one set the the proxy's IP and port number in the \emph{prefs.js} file in the profile folder. 
In Chromium browsers, custom root CAs can be added to the \code{.pki/nssdb/} folder located either under the user's home folder and trusted by all browsers or created under individual profile folders when browsers are installed with \flatpak. 
Section~\ref{sec:attacks_mitms} demonstrates how we leverage the \emph{certutil} utility to add custom root CAs to Chromium and Firefox browser profiles and mount MiTM attacks~\cite{CertUtil}.

\subsection{Other User Data}
\label{sec:browsing_profiles_history}
Various files, e.g., \code{places.sqlite} on Firefox or \code{History} for Chromium, are used to keep track of the user's browsing history, most visited websites, bookmarked pages, downloads, autofill, and search information, etc. These files do not enjoy any particular security mechanisms on the user's disk. Therefore, they are readily accessible. While unprotected, most of this user data is privacy-sensitive. For example, autofill data may contain personal information such as people's names, addresses, or emails. 
Beyond custom proxy settings on Firefox, the \code{prefs.js} file can contain many other entries. Notably, Safebrowsing~\cite{GoogleSafeBrowsing} checks can be influenced by manipulating the options of the form \code{browser.safebrowsing.*}.

%% file: fsa_api.tex
\section{File System Access API}
\label{sec:fsa}
\label{sec:file_system_access_api}
\label{sec:fsa_vs_browser_profiles}

\done{Add a Figure with 3 different steps to illustrate everything}
\begin{figure*}
    \minipage{0.33\textwidth}
     \begin{subfigure}{0.98\linewidth}
        \centering
        \includegraphics[width=0.99\linewidth]{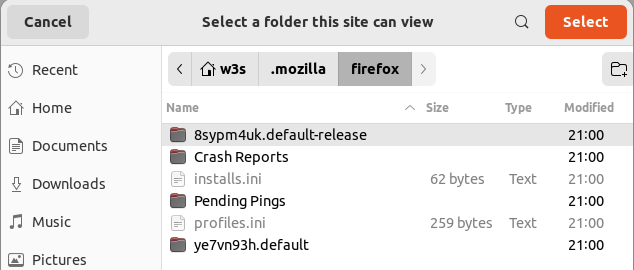} 
        \caption{File Browsing}
        \label{fig:fsa_file_browsing}
     \end{subfigure}
    \endminipage
    \minipage{0.33\textwidth}
     \begin{subfigure}{0.98\linewidth}
        \includegraphics[width=0.99\linewidth]{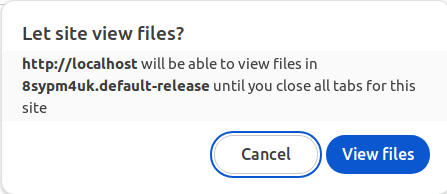} 
        \caption{Read Permission}
        \label{fig:fsa_read_permission}
     \end{subfigure}
     \endminipage
     \minipage{0.33\textwidth}
     \begin{subfigure}{0.98\linewidth}
        \includegraphics[width=0.99\linewidth]{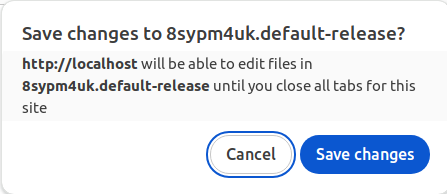} 
        \caption{Write Permission}
        \label{fig:fsa_write_permission}
     \end{subfigure}
     \endminipage\hfill
    \caption{File System Access API working: browsing, granting read and write permission}
    \label{fig:fsa_browsing_read_write}
\end{figure*}

The File System Access (FSA) API is a modern JavaScript API supported by Chromium browsers for interacting with the user's physical file system. As we later discuss in this paper, attackers might attempt to use this API to compromise profile folders. So, below, we present its functionality and associated security mechanisms.
Traditionally, websites have interacted with users' file systems via different mechanisms combined. Notably, the \code{<input type=file [webkitdirectory]>} HTML snippet can be used to trigger a file(s) picker dialog that lets the user browse the file system in search of specific files or directories. The file(s) can then be uploaded to a remote server, read, manipulated, and the result written back to the file system with an HTML snippet like \code{<a download="...">}. 
With the FSA API, all these operations come in handy to websites via JavaScript. The main steps are summarized in Figure~\ref{fig:fsa_browsing_read_write}. The \code{showDirectoryPicker()} or \code{showOpenFilePicker()} global functions trigger file system browsing, returning a file or directory handle object to the calling program. The handle serves to interact with the file system, read its content, and write modifications back on disk by invoking the \code{showSaveFilePicker()} global function. 
%
%
%
\\
{\bf Security mechanisms.} 
Multiple defense-in-depth security mechanisms guard the FSA API. In addition to requiring user interaction to browse the file system, the browser checks selected folders against a blocklist and further check files against GoogleSafeBrowsing~\cite{GoogleSafeBrowsing} and tag them with the  mark-of-the-web~\cite{Mark_of_the_Web}, the latter being used by operating systems to warn users that a binary they are about to run was obtained from the web and therefore, may be unsafe. 
\\
{\bf The blocklist}. It lists folders and files inaccessible via the FSA API~\cite{FSA_Blocklist}. This includes notably sensitive system folders like \code{/etc, /boot, /dev, /proc, /sys} on Linux\footnote{And similar folders on other operating systems} or the user's home directory (e.g., \code{\$HOME}). Note that if access is denied to the user's home directory root, subfolders that it contains may be accessed, provided that they are not blocklisted (e.g., \code{\$HOME/.config, \$HOME/.dbus, \$HOME/.ssh} are forbidden). Folders accessible by default include \emph{Desktop, Downloads, Documents, Music, Pictures, Videos}, etc. Notable blocklist entries are Chromium browser profile folders (See Section~\ref{sec:browser_profiles}), and access to the browser's own profile folder is forbidden.

%% file: threatmodel.tex
\section{Threat Models}
\label{sec:threatmodel}
\label{sec:threatmodels}
In this work, we propose different file system attackers that can compromise browsers by violating the confidentiality and integrity of profile folders. 
%
Considering the trove of sensitive data contained in browser profiles, we believe that, as part of the defense-in-depth strategy, \textbf{browsers should assume that the file system is untrusted and that attackers can attempt to both read and write files stored in profiles that are persisted on the disk}. Inspired by prior work on secure password storage~\cite{NaiakshinaDGZ019,AcarSWMF17}, 
we assume that attackers can access and alter the information stored on the disk. Below, we define multiple file system attackers that can achieve these malicious goals using various capabilities. 

%
We first describe local attackers that can access the file system with the capabilities of a \emph{non-privileged} user, as opposed to a superuser. The proposed attackers can only \emph{read}~\sx{sec:fs-read}, read and \emph{write}~\sx{sec:path_traversal_attacker} files in profile folder, or even \emph{execute} local commands~\sx{sec:supply_chain_attacker}. We present these local attackers starting from weak to strong ($A_R<A_W<A_X$), and we note that they are all weaker than a malware attacker. We illustrate concrete instantiations of attacks in each section and argue about their feasibility. Finally, we discuss a weaker web attacker in Section~\ref{sec:fsaapi}. We further argue about the utility of the proposed threat models in Section~\ref{sec:feasibility}.



\subsection{Local File System \emph{Read Attacker} ($A_R$)}
\label{sec:fs-read}

This attacker can only read files from browsers' profile folders. They cannot modify files or execute them as programs. However, they can send data to attacker-controlled machines. As discussed in Section~\ref{sec:compromizing_browser_profiles}, this attacker can perform powerful read-only attacks like session hijacking or history sniffing.
Chromium browsers partially consider this attacker model by encrypting user passwords and cookies before storing them in the profile folder.
Below, we describe two instantiations of this attacker.
\\
{\bf Shared Linux machine.}
Surprisingly, by default, most modern Linux distributions allow unrestricted read and execute access to all files in the home folder of all users\footnote{\url{https://ubuntu.com/server/docs/user-management\#user-profile-security}}. Unless this default behavior is overwritten, in a multi-user setup, untrusted users can exfiltrate any file in the home directory of other system users. Since browser profiles are often stored in home folders on Linux, this can enable powerful profile compromise attacks. 
%
\\
{\bf Self-exfiltration of profiles.} Attackers might attempt to lure the victims into sending them the content of their profile folder directly, e.g., by promising financial rewards or benign functionality such as cross-browser session synchronization, as proposed by the defunct Xmarks~\cite{Xmarks} browser extension. This is analogous to self-XSS~\cite{SelfXSS}. 


\subsection{Local File System \emph{Write Attacker} ($A_W$)}
\label{sec:path_traversal_attacker}
This attacker can read and write files in browser profile folders. We do not assume the attacker can modify system binaries, such as the browser itself.
This attacker can mount most attacks in this work, which we deem significantly weaker than a malware attacker. The attacker only has file system access to the profile folder but no code execution privileges. We note that Chromium browsers partially consider this attacker when attempting to detect unlawful modifications of browser extensions stored on the disk.
\\
{\bf Path traversal.}
Arbitrary file writes are usually discussed in the realm of remote web servers~\cite{PathTraversal-OWASP} where a benign-but-buggy web service (or web server) is abused to access the file system on which it executes. We believe it is feasible for users to operate such benign but buggy programs, e.g., a locally run printing server or an antivirus web UI service. This program could have been installed by the user deliberately, or the attacker leveraged social engineering techniques to have the user install and run it. Then, when the user visits a malicious website under the attacker's control, the latter can start scanning the user machine, i.e., \code{http://localhost/}, in search of vulnerable web services. The likelihood and success of this attacker depend on the level of vulnerabilities (read, write, or execute capabilities) exhibited by the buggy program. 
Oz et al.~\cite{OzAATKU24} show that unrestricted file uploads are prevalent in real-world web applications, allowing attackers to write arbitrary files on the file system.
Another variant of this attack is to exploit Zip Slip, a type of path traversal vulnerability reported in archive programs by Snyk in 2018~\cite{ZipSlip}, where an attacker crafts a malicious zip file containing links pointing to arbitrary upper locations relative to where the user will decompress the archive. Assuming that the victim stores and decompresses the malicious zip in the Downloads folder (generally located under the user's home folder), then the attacker 
can overwrite many sensitive folders during decompression like browser profiles (e.g., \code{../.config/google-chrome/} and \code{../.mozilla/firefox/} for the Chrome and Firefox default folders on Linux). Using such payloads, they can compromise the integrity of browser profiles, as we discuss in Section~\ref{sec:compromizing_browser_profiles}.

\subsection{Local File System \emph{Execute Attacker} ($A_X$)}
\label{sec:supply_chain_attacker}

This attacker can read and write files in a profile folder and execute operating system commands. However, they cannot necessarily add or modify such commands, e.g., by adding execution rights to a file or modifying the browser's binary. While malicious code can act as an $A_X$ attacker, the malware attacker is usually more powerful and capable of arbitrary system changes. Nonetheless, different Google teams report the existence of two classes of relevant malware prevalent in the wild: cookie-theft~\cite{cookietheftmalware} and infostealing~\cite{infostealingmalware} malware. Browser profiles are a common target for this class of unwanted software, and vendors adopt mechanisms such as encrypting cookies to hinder the success of such attacks. As mentioned in the introduction, there are multiple reports of NPM malware targeting browser profiles~\cite{steal-passwords,steal-cookies,steal-profiles,steal-ext-data}. All this empirical evidence suggests that $A_X$ attackers are common in the wild, and browser vendors are already taking actions~\cite{App-bound-enc-Windows} to hinder their success rate.
\\
{\bf Supply chain attacker or malware}
This attacker tricks the user into installing and running malicious code on their device by exploiting third-party dependencies~\cite{DuanBJAXISL19,LadisaPMB23}, e.g., NPM packages or PIP modules. While this is a strong attacker, we note an increase in the prevalence of such attacks in recent years. While the consequences of supply chain attacks are disastrous, i.e., arbitrary code execution, we argue that attackers often aim for financial gains, e.g., stealing crypto wallets~\cite{steal-ext-data} or access tokens~\cite{cookietheftmalware}. Hence, a man-in-the-browser attack as described in Section~\ref{sec:attacks_extensions} would allow them to hijack financial transactions, impersonate users, or bypass two-factor authentication; all this by using convenient, easy-to-use web technologies instead of having to deploy malicious browser binaries. 
Given their catastrophic consequences and prevalence, vendors should consider these attacks seriously.


\subsection{Remote Web Attacker ($A_{Web}$)}
\label{sec:fsaapi}

This attacker aims to use browser-specific APIs to alter profile folders on the disk from inside a web browser. They either attempt to access the underlying browser's profile folder or another browser in a multiple-browser setup, i.e., when the user installs multiple browsers. Moreover, we assume that the attackers can execute malicious code in the browser in the first place, e.g., by convincing the user to visit a website they control via a spear phishing campaign. With this attacker model, we aim to study whether web attackers can become file system attackers and under which circumstances. 
\\
{\bf The File System Access (FSA) API}
 As discussed in Section~\ref{sec:fsa}, the FSA API is built on two security pillars: it has a list of blocked folders referred to as the \emph{blocklist} that are inaccessible by default, and it requires that users browse the files and folders that are to be accessed and grant explicit read and write permissions. Hence, the remote attacker needs to bypass the blocklist and obtain consent from the user to modify the files corresponding to the profile. In Section~\ref{sec:fsa_vs_browser_profiles}, we leverage shortcomings in the specification and implementation of this API to demonstrate profile compromise attacks against major browsers performed by remote attackers. We note that obtaining user consent as part of the attack limits the attack's scalability, requiring 
 social engineering tactics to deceive the user. 
 Its devastating consequences, i.e., profile compromise, make it worth considering. Moreover, prior work by Oz et al.~\cite{Oz-etal-23-USENIX} uses similar assumptions in the threat model. Finally, we also perform a user study~\sx{sec:userstudy} to show the feasibility of the proposed attack.

\subsection{Utility of the Proposed Threat Model}
\label{sec:feasibility}
%
%
{\bf Vendors' attitude towards 
the file system attacker.}
We observe that browser vendors' attitudes toward file system attackers are contradictory and continuously changing. While both the Chrome Security FAQ~\cite{chromesecfaq} and the developers responding to our disclosures~\sx{sec:responsible_disclosure} state that such attackers are outside Chrome's threat model, they deploy security controls~\sx{sec:background} to protect against such attackers, which they also deem to be prevalent in the wild~\cite{cookietheftmalware,infostealingmalware}. Recently, Chrome announced additional security controls ~\cite{DBSC-Chromium,App-bound-enc-Windows} to be deployed to detect file system attackers. While Firefox is falling behind with adopting similar mechanisms, its security team is very keen on considering this attacker~\sx{sec:responsible_disclosure} shortly. Thus, researchers must investigate this threat to model browsers and inform the vendors in their decision-making.
%
%
\\
{\bf Large-scale vs. targeted attacks.}While powerful, none of the file system attackers described earlier can mount a large-scale profile compromise attack on billions of web users. 
Instead, we believe they are more adequate for targeted attacks, in which the adversary and the victim share the same machine or the attack requires multiple user actions, like in the case of FSA API attacks in Section~\ref{sec:fsa_vs_browser_profiles}. We believe that such targeted attacks are relevant, especially when dealing with high-stake targets~\cite{StaicuP19,BlondUGCSK14,ho2017detecting}, whose compromise might lead to financial gains or unlawful private information leakage, e.g., disclose private photos of celebrities from their cloud storage accounts.
A recent study surveying various social tactics concluded that social engineering will be one of the most prominent challenges of the upcoming decade~\cite{Salahdine-Kaabouch-19-FI}. At the same time, the European Union Agency for Cybersecurity suggests that most cyberattacks nowadays include some form of social engineering~\cite{SocialEngineeringENISA}. 
Thus, targeted attacks like the ones considered in this work are worth studying, especially when taking into consideration their potentially catastrophic impact, e.g., full browser compromise, 
and the fact that weak web adversaries can mount some of them.
Thus, we believe that browser vendors should consider the file system attacker as a serious threat and deploy effective solutions to protect sensitive user data stored on the disk. Intriguingly, profile compromises are made possible by legitimate features of modern browsers, like the portability of profiles stored on the disk and the real-time synchronization of profiles across multiple user machines, which might amplify the effect of an attack.

%% file: fs_vs_browser_profiles.tex
\begin{table}
    \footnotesize{
    \centering
    \begin{tabular}{|p{3cm}|p{0.17cm}|p{0.24cm}|p{0.18cm}|p{0.17cm}|p{0.24cm}|p{0.18cm}|p{0.17cm}|p{0.24cm}|p{0.18cm}|p{0.15cm}|p{0.15cm}|p{0.15cm}|p{0.15cm}|p{0.15cm}|p{0.15cm}|p{0.15cm}}
        \hline
        \bfseries{Attacks} & \multicolumn{3}{c|}{\bf Firefox } &  \multicolumn{3}{c|}{\bf Chromium} &  \multicolumn{3}{c|}{\bf Safari} \\ \hline 
         \bfseries{Attacker type} & \bfseries{$A_R$} & \bfseries{$A_W$} & \bfseries{$A_X$} & \bfseries{$A_R$} & \bfseries{$A_W$} & \bfseries{$A_X$} &
         \bfseries{$A_R$} & \bfseries{$A_W$} & \bfseries{$A_X$}
         \\ \hline 
         \rowcolor{lightgray}Session hijacking & \greencheck & \greencheck  & \greencheck & & \greencheck & \greencheck  & & & \\ \hline 
          \rowcolor{gray}MiTM attack & & \greencheck  & \greencheck & & \greencheck & \greencheck  & & & \\ \hline 
         \rowcolor{lightgray}Installing malicious extensions   & & \greencheck & \greencheck & & \greencheck & \greencheck  & & & \\ \hline
          \rowcolor{gray}History sniffing & \greencheck & \greencheck & \greencheck & \greencheck & \greencheck & \greencheck & \greencheck & \greencheck & \greencheck  \\ \hline 
         \rowcolor{lightgray}Password  exfiltration & & \greencheck & \greencheck & & \greencheck & \greencheck & & &  \\ \hline 
          \rowcolor{gray}Preference changes  & & \greencheck & \greencheck & & \greencheck & \greencheck  & & &  \\ \hline 
        \rowcolor{lightgray}Devices permission alteration  & & \greencheck & \greencheck & & \greencheck & \greencheck  & & \greencheck & \greencheck \\ \hline 
          \rowcolor{gray}Exfiltration of auto-fill data & \greencheck & \greencheck & \greencheck & \greencheck & \greencheck & \greencheck & & &  \\ \hline 
    \end{tabular}
    \caption{Profile compromise attacks that can be mounted against web browsers by the considered attackers. $A_R$ can only read, $A_W$ can also write, and $A_X$ execute files on disk. 
    }
    \label{tab:security_attacks_capabilities}
    }
\end{table}

\section{Compromising Browser Profiles}
\label{sec:compromizing_browser_profiles}
\label{sec:fs_vs_browser_profiles}

Provided a file system attacker with different capabilities, we summarize the various attacks that can be performed against browser profiles. Table~\ref{tab:security_attacks_capabilities} shows an excerpt of attacks. The following subsections provide additional details for many of these attacks. 

\subsection{Bypassing Chromium Cookies Encryption}
\label{sec:attacks_cookies}
A natural approach to bypassing encryption would be to engage in the process of reverse-engineering the key used by the browser to perform the cryptographic operations, as prior work has attempted~\cite{Sanchez-etal-20-CANS}. Instead, we found that, even without knowing the key, the encryption is trivially bypassed by an attacker with write capability. The critical observation we made by analyzing the way the encrypted values were stored in the profile folders is that, in all cases, there is not a hard link between the encrypted values and the website they belong to. In other words, the messages are not \emph{authenticated}~\cite{HMAC}. Therefore, an attacker can rewrite or copy the cookies in the database and assign them to a website under their control. Then, when the user visits this website, the browser will decrypt the cookies and make them readable to the attacker. 
%
Listing~\ref{lst:chromium_cookies_encryption_bypass} demonstrates this process where cookies from \url{youtube.com} are copied and assigned to \url{attacker.com}
%
%
\begin{lstlisting}[basicstyle=\tiny\ttfamily,language=SQL, caption={SQL query to bypass Chromium encryption of cookies. All the cookies of the target website, youtube.com are copied and set to the attacker domain attacker.com.}, label={lst:chromium_cookies_encryption_bypass}]
INSERT INTO cookies (host_key,top_frame_site_key,value,is_httponly,creation_utc,name,encrypted_value,path,expires_utc,is_secure,last_access_utc,has_expires,is_persistent,priority,samesite,source_scheme,source_port,is_same_party,last_update_utc) 
SELECT ".attacker.com","","",0,creation_utc,name,encrypted_value,path,expires_utc,is_secure,last_access_utc,has_expires,is_persistent,priority,samesite,source_scheme,source_port,is_same_party,last_update_utc FROM cookies
WHERE host_key = "youtube.com";
\end{lstlisting}
%
%
Once the browser launches and the user visits the attacker's website, the cookies are automatically decrypted and sent to the attacker. To force such visits, attackers can modify the browser's homepage to point to their website; thus, the cookies will be automatically sent when the browser restarts. 
As one can notice, the \code{HttpOnly} flag is also disabled, making the cookies directly accessible to JavaScript. The read cookies can be used to populate a browsing session under the attacker's control, leading to the hijacking of the user's browsing session~\cite{SessionHijacking}. The consequences can range from financial loss to access to sensitive data like photos or emails. 

\subsection{Bypassing Passwords Encryption}
\label{sec:attacks_saved_credentials_and_passwords}

The intuition about bypassing the encryption process on the saved credentials, even for Firefox, is the same as described for encrypted cookies in Chromium browsers (Section~\ref{sec:attacks_cookies}): the attacker makes copies of the encrypted passwords, changes the site name to one under their control and update the original databases with the new entries. 
Listing~\ref{lst:chromium_passwords_encryption_bypass} shows a code snippet used to steal the user's Facebook credentials in Chromium. 

\begin{minipage}[center]{0.48\textwidth}
\begin{lstlisting}[basicstyle=\tiny\ttfamily,language=SQL,caption={SQL query to bypass Chromium encryption of passwords. All the credentials of the target website, i.e. \url{facebook.com} are copied and set to \url{attacker.com}. To retrive the stolen credentials, the attacker adds a form to their home page with input fields named "email" and "pass".}, label={lst:chromium_passwords_encryption_bypass}]
INSERT INTO logins (origin_url,action_url,username_element,username_value,password_element,password_value,signon_realm,date_created,blacklisted_by_user,scheme,password_type,times_used,form_data,skip_zero_click,generation_upload_status,date_last_used,date_password_modified,date_received,sharing_notification_displayed) 
SELECT "https://attacker.com", "https://attacker.com/", "email", username_value, "pass", password_value, "https://attacker.com/", date_created,blacklisted_by_user,scheme,password_type,times_used,form_data,skip_zero_click,generation_upload_status,date_last_used,date_password_modified,date_received,sharing_notification_displayed FROM logins 
WHERE origin_url = "https://www.facebook.com/"
\end{lstlisting}
\end{minipage}

Once the user visits the website under the attacker's control, the home page has a form, which the browser will automatically populate the stolen credentials after decrypting them. The attacker can now read and leverage the credentials in the backend to log in as the user~\cite{SessionHijacking}. 
To force visits to attacker-controlled websites, attackers can again modify the browser's homepage and deploy phishing-like strategies on their website to trick the user into submitting the form.
A similar process was followed for stealing Firefox browser credentials. Therefore, we omit further details. 

\subsection{Browser Extensions and Code Executions}
\label{sec:attacks_extensions}
\label{sec:infecting_chromium_extensions_fsa}

For adversaries, extensions represent an interesting target because, in addition to code execution (i.e., XSS~\cite{OWASP_Types_Of_XSS}), the attacker can abuse the extension's privileges and its powerful APIs to access any user information: content scripts can be injected in web pages to mount keylogging attacks and read login credentials; the background script can inspect/modify HTTP requests and responses, etc. 
%
%
\\
{\bf Adding a signed malicious extension to Firefox.} 
Instead of trying to bypass the integrity checks performed by Firefox, the intuition about this attack is rather to develop and upload a malicious-but-signed extension to a repository that Firefox trusts, e.g., AMO~\cite{AMO-Firefox}, then install it in a Firefox profile, and copy the metadata generated by the browser and the extension signed XPI bundle to the victim user's browsing profile. Specifically, two (2) files are to be copied:
(i) \code{extensions.json}, which contains the list of installed extensions and the extension XPI bundle, which is stored in the \code{extensions/} folder. 
Once the user launches their browser, the malicious extension(s) will be loaded and executed without them noticing. The browser will not complain because Mozilla signed the extension. 
Malicious extensions are regularly found on AMO~\cite{AMO-Malicious-Addons}. Malicious extensions can also be kept unlisted so researchers scanning these repositories do not find and report them, potentially leading to removal. Finally, an attacker can host an extension on a server under their control as long it is served with the correct \code{application/x-xpinstall} MIME type.
We also follow similar steps to install extensions in \emph{Developer mode} in Chromium by modifying \texttt{Preferences} while carefully preserving the integrity protection values~\cite{Sanchez-etal-20-CANS}. 

\noindent
{\bf Infecting a signed Chromium extension.} 
From discussions with various vendors, it appeared that the integrity checks on extensions are not enabled by default in the Chromium sources. It is the responsibility and choice of each vendor to enable it. When one builds a Chromium browser from the sources or simply chooses not to enable this feature, signed extensions that the user will install can be manipulated. This affects the raw Chromium build and all users building their own browsers from the sources, e.g., to disable Google Services for privacy reasons~\cite{Ungoogled-Chromium}. More interestingly, among the official browsers we have tested, we found that Vivaldi does not enable integrity checks on extensions.  As the company claims almost 3.1 million active users for its cross-platform browser~\cite{VivaldiUsers}, the portion of exposed (desktop) users may be rather substantial. All Chromium browsers fail to perform integrity checks properly when considering an execute-able attacker.  We refer the interested reader to the Appendix, which gives more details about the issue. 


\subsection{Altering Users Permissions}
\label{sec:attacks_permissions}

Listing~\ref{lst:attack_permissions_firefox} shows how to modify the \code{permissions.sqlite} database in a Firefox profile to grant the attacker access to the camera.  
%
\begin{lstlisting}[basicstyle=\tiny\ttfamily,language=SQL,label={lst:attack_permissions_firefox}, caption={SQLite query for editing the \code{permissions.sqlite} database to grant the attacker access to the user's camera.}]
INSERT INTO moz_perms (origin, type, permission, expireType, expireTime, modificationTime) values ("https://attacker.com", "camera", 1, 0, 0, 1702964711537);
\end{lstlisting}
%
Access to the microphone, GPS, or the ability to send them web push notifications is granted similarly.
%
%
Concretely, when the user visits the attacker's site, the browser will not prompt the user to grant permission to access these features as they would have been automatically and silently granted to the malicious website. The attacker can then turn on the user's microphone or camera and track their physical location, and with a registered service worker, they can serve web push notifications~\cite{WebPushNotifications}. In Chromium browsers, the manipulations are similar and therefore omitted. 
%
%

\subsection{Traffic Interception Attacks}
\label{sec:attacks_mitms}
To mount a Man-in-the-middle (MiTM) attack where all requests and responses are redirected to an HTTPS proxy, one has to be able to (i) make the browser trust a custom root certificate authority (CA) and (ii) additionally instruct the browser to redirect all the traffic to the MiTM proxy server. The \code{certutil} utility from the Network Security Service libraries~\cite{NSSDB} makes it handy to add custom CAs to certificate databases in browser folders. Listing~\ref{lst:attack_mitm_proxy} shows an example. 
\begin{lstlisting}[basicstyle=\tiny\ttfamily,label={lst:attack_mitm_proxy}, caption={Adding a custom CA to a CAs database: \code{profile\_folder} is a folder containing the \code{cert9.db} and \code{key4.db} files and \code{mitmproxy-ca-cert.pem} is the malicious root CA to be tagged \code{mitmproxy} in the CAs database.}, language=bash]
certutil -d profile_folder -A -t C -n mitmproxy -i mitmproxy-ca-cert.pem
\end{lstlisting}
The attacker is left to run a MiTM server and indicate its location (IP, port) by editing the browser profile network proxy settings.
\\
{\bf Firefox browsers proxy settings.}
As shown in Listing~\ref{lst:attack_mitmproxy_ip_port_firefox}, the MiTM server's location (IP and port) is set in the \code{prefs.js} file. 
\begin{lstlisting}[basicstyle=\tiny\ttfamily, language=JavaScript, label={lst:attack_mitmproxy_ip_port_firefox}, caption={Indicating the location (IP and port) of the MiTM proxy to Firefox. Here we use a private local IP address, but any (publicly) accessible IP can be specified instead.}]
user_pref('network.proxy.http', '127.0.0.1');
user_pref('network.proxy.http_port', 8080);
user_pref('network.proxy.ssl', '127.0.0.1');
user_pref('network.proxy.ssl_port', 8080);
user_pref('network.proxy.type', 1);
\end{lstlisting}
These settings are specific to the browser profile they apply to. They do not affect other Firefox browsers or interfere with network proxy settings at the operating system level.
\\
{\bf Chromium browsers.}
In most cases, Chromium browsers do not write proxy settings in the profile folder but rely on the OS-level static proxy settings or allow dynamically specifying the settings with the \code{----proxy-server} command line or by using the \emph{proxy} API of webExtensions. With the latter, the attacker tricks the user into installing a minimal browser extension. This extension's logic consists of invoking the proxy API to indicate the location (IP and port) of the HTTPS proxy server~\cite{WebExtensionProxyAPI}. We developed, published, and kept unlisted such an extension on CWS~\cite{XPROXY} for our experiments.
\\
{\bf Impact of a MiTM attack against browsers.}
When the user launches the infected browser, all of their traffic is silently redirected to the MiTM proxy, where all HTTPS content is viewed in the clear and can be manipulated to inject additional code, log and exfiltrate credentials, hijack financial transactions, etc. This attack is potent because even requests and responses issued by the browser, e.g., to check for updates or to download, verify, and install extensions, can be tampered with. 
We extensively used the Mitmproxy~\cite{Mitmproxy} proxy in our demos and developed a Python addon to view and manipulate request and response headers and bodies.

%% file: fsa_vs_browser_profiles.tex
\begin{figure}
    \centering
    \includegraphics[width=0.80\linewidth]{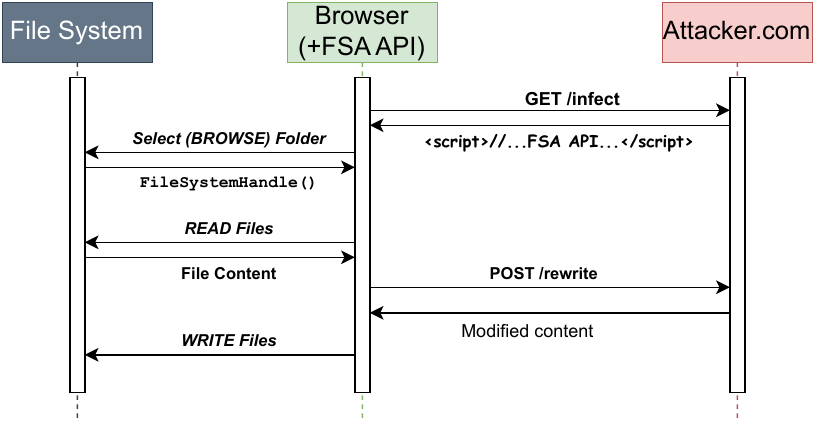}
    \caption{Attack setup to read and write browser profile folders on the user's file system with the FSA API}
    \label{fig:fsa_attack}
\end{figure}

\subsection{Remote Web Attacks}
\label{sec:fsa_vs_browser_profiles}
We assume an $A_{Web}$ attacker that aims to use the FSA API to compromise browser profiles. We first discuss the setup, our testing methodology, and the findings below.
%
%
\\
{\bf Setup.}
\label{sec:fsa_attacks_setup}
Figure~\ref{fig:fsa_attack} depicts the typical attack and testing workflow. Files in browser profile folders are made of numerous text and binary formats. Manipulation of text formats, e.g., JSON data and JavaScript code, can be done directly within the attacker's website. For more evolved content types, schemas, and formats like SQLite, certificate databases, and binaries in general, the choice we have made resolves around reading the content from the file system, serializing, and sending it to a back-end server where the content can be easily manipulated using tools like the SQLite NPM module~\cite{SQLite3NPM}, or the \code{certutil} utility~\cite{NSSDB}. The resulting modifications are then downloaded and written back to the user's file system.
\\
{\bf Testing strategy.} 
To identify ways of misusing the FSA API for profile compromise, we systematically study its specification to identify controls intended to prevent abuses. We identify the blocklist~\cite{FSAAPI-Block} as the primary mechanism. It contains eight clauses. We defined one test case for each, which we manually executed in all the considered browsers and operating systems (See Table~\ref{tab:browser_default_profile_folders}). For the specification that agents should restrict access to ``directories where the user agent stores website storage'', our test checks all the known folders where browsers usually store data on the disk (See Table~\ref{tab:study_desktop_browsers} in the appendix). For OS-dependant specifications like the ones involving DLLs or \texttt{.lnk} files, we generalize them to their equivalent in other operating systems, e.g., \texttt{.so} files and symbolic links on Linux. 
%
%
Contrary to prior work by Oz et al.~\cite{Oz-etal-23-USENIX}, which concluded that browser profiles are inaccessible using FSA API, we identify concrete ways to abuse this API for remote compromise. We discuss two high-level methods below:
\\
{\bf Profile folders missed by the blocklist.}
\label{sec:threats_blocklisting_missing_folders}
First, we observed that Gecko-browsers profile folders are entirely ignored by the FSA blocklist when these browsers are regularly installed by the underlying Linux system package manager, e.g., \code{apt} or \code{dnf}. This is the case for Firefox, whose profile folders are located under \code{.mozilla/} in the user's home directory, but also Waterfox (\code{.waterfox/}) or LibreWolf (\code{.librewolf/}) browsers. \footnote{Tor is also in this case, but this browser stores minimal information on the disk, and therefore we exclude it}. 
Figure~\ref{fig:fsa_browsing_read_write} demonstrates the simple steps followed by the user to grant access to a Firefox profile folder. 
Furthermore, the blocklist does not consider browsers installed via package managers like \code{snap}~\cite{SNAPD} or \code{flatpak}~\cite{Flatpak}. Indeed, the \code{Snap} utility stores the profile folders of browsers it manages in the \code{snap/} folder under the user home directory. This includes Firefox and Chromium browsers like Opera or Brave. Likewise, the \code{flatpak} utility stores browser profile folders in the \code{.var/} folder in the user's home directory. This utility case is worrisome because it bundles most major browsers, including Chrome, Edge, and Firefox. 
These shortcomings of the blocklist enable cross-browser compromises, in which a Chromium browser is used to attack other browsers. 
%
\\
{\bf Symbolic links.}
\label{sec: threats_symbolic_links}
    In this scenario, the attacker creates a symbolic link pointing to the file system's root on a machine under their control. Note that the command does not require root privileges to succeed. Then, the folder is compressed to preserve the symbolic link. The resulting zip file is only a few bytes long—173B in our demos—because no actual content is copied into the ZIP file. 
%
The attacker then tricks the victim into downloading and unzipping the file in a location where it is accessible with the FSA API, e.g., the user's home directory, the \code{Desktop}, or more simply, the \code{Downloads} folder where most browsers store content downloaded from the Web. The \code{Downloads} folder is interesting because even restricted systems like macOS' make it accessible by default to most applications, including web browsers. The final steps are similar to the ones described in Figure~\ref{fig:fsa_browsing_read_write}:  the user selects and grants the attacker's website access to the folder they have just unzipped or to one of its parents. Once completed, the attack allows unrestricted access to all the files on the disk, including the ones on the blocklist. This attack affects all (Chromium) browsers on UNIX operating systems. 

While the issues above might be considered expected bugs in a newly introduced API, we believe they show something more significant. As discussed in Section~\ref{sec:fsaapi}, the introduction of the FSA API fundamentally changes the web security model, allowing web attackers to become file system attackers. On the one hand, the Chromium team is adopting a pragmatic blocklist solution~\cite{Chromium-issue-discussion} to limit blatant misuse; on the other hand, they do not consider bugs in this security mechanism a security bug~\cite{chromesecfaq}, per se, arguing that user consent is hard to obtain. In a user study~\sx{sec:userstudy}, however, we show that users carelessly grant read/write consent to websites, underestimating the dangers they expose themselves to. We believe that the skepticism of other vendors to adopt this API is justified, and we encouraged both the Chromium team and the research community to critically reassess whether this fundamental departure from the traditional web attacker model is worth the associated risk, as advocated by Snyder et al.~\cite{SnyderTK17}.


\subsection{Case Studies}
\label{sec:fsa_attacks_case_studies}
We mounted and validated all the attacks discussed in this work (Section~\ref{sec:attacks_cookies} to~\ref{sec:fsa_vs_browser_profiles}). We refer the reader to demos~\cite{PoC} that we have recorded and the materials for the attacks. Examples of demos include MiTM attacks, silently granting permissions to use the camera or microphone or send web push notifications, bypassing the encryption and exfiltrating cookies and passwords, bypassing the integrity checks on browser extensions, etc. We also include a demo of the proof-of-concept defense, which consists of encrypting the browser profiles to prevent manipulations by the attacker. 

%% file: ustudy.tex
\begin{table}[!htp]
    \footnotesize{
    \centering
    \begin{tabular}{|c|c|}
        \hline
            & \bfseries{\#Participants} \\ \hline 
         \multicolumn{2}{|c|}{Questionnaire download mode (N=\participantsfinal)} \\ \hline
         \rowcolor{lightgray}\textit{ZIP} & \downloadzip \\ \hline 
         \rowcolor{gray}\textit{FSA} & \downloadfsa \\ \hline 
         \multicolumn{2}{|c|}{Responses upload mode (N=\participantsfinal)} \\ \hline
          \rowcolor{gray}\textit{FSA} & \uploadfsa \\ \hline 
         \rowcolor{lightgray}\textit{Traditional} & \uploadtraditional \\ \hline \hline 
          \rowcolor{gray}\textbf{ZIP (download) + FSA (upload)} & \downloadzipuploadfsa \\ \hline 
    \end{tabular}
    \caption{Summary of user study's findings.}
    \label{tab:userstudyresultsumary}
    }
\end{table}

\subsection{User Study}
\label{sec:userstudy}
We report on the results of a user study that assessed users' susceptibility to social engineering with the FSA API. More precisely, we set out to evaluate if users would follow the steps of the symbolic link attack described earlier and give access to their entire disk. In particular, under the cover of responding to a questionnaire (that can be found at the end of the appendix), participants are invited to visit a website, download an archive, unzip it, and fill out a questionnaire (presented in an HTML file), save their responses as a PDF, and finally upload the PDF by indicating the folder where it is stored. This final step can be achieved with a traditional upload dialog or with the FSA API, where the participants are prompted with the different permissions dialogs to grant access to their file system. 
As described in Section~\ref{sec:fsa_vs_browser_profiles}, an unethical attacker could include a symbolic link in the ZIP file, which we did not do in this study. Instead, we measure how many users complete the task by downloading via ZIP and uploading with the FSA API, i.e., the steps of our attack, and thus, allow access to their entire file system.
\\
{\bf Setup and ethical considerations.}
Though not the primary focus of our work, 
we designed the study to abide by the highest ethical and privacy preservation principles. 
We first performed a preliminary internal evaluation to test and improve the questionnaire and the setup. We then informed the ethics team at our institute about our study's design. The team found nothing unethical or violating users' privacy since we do not collect private or personal information about participants. Finally, we run the survey on the Prolific platform~\cite{Prolific}. In particular, we request that candidate participants are fluent in English and willing to take the study in a desktop browser. We could not specify that participants should use a Chromium browser on Prolific, but we did it on the study's website. We also disclosed that participants will have to download resources on their devices. We extensively refined our study to comply with Prolific's requirements, e.g., on the website's landing page, participants are informed about the study, the steps they will follow, and the approximate time it would take to complete it (5-10 minutes). 
To proceed, they must read and review the consent form and freely consent to participate in the study. The participants were in control and could quit the study and withdraw at any moment. 
Finally, users who consented provide their Prolific ID and are redirected to the instructions page, where the study is completed in three (3) steps, with multiple options:
%
%
%
\begin{enumerate}
    \item {\bf Downloading the questionnaire:}
        \begin{itemize}
            \item         
        {\bf ZIP}: The participant downloads and manually unzips the file to get access to the questionnaire.
        \item
        {\bf FSA API}: The participant is asked to select a directory where the questionnaire will be automatically written via JavaScript using the FSA API. The user is prompted to grant read permission to access their physical FS.
        \item
        {\bf Git}: The participant manually clones a GitLab repository containing the questionnaire
        \end{itemize}
    \item {\bf Fill in the questionnaire}: The participants fill in the local HTML file and save their responses as PDFs.
    \item {\bf Upload the responses:}
        \begin{itemize}
            \item         
       {\bf FSA API}: First, the participants drag and drop or browse the responses' folder. Then, the PDF with the responses is discovered and uploaded automatically. Subsequently, the user is prompted to grant permission to interact with the file system. 
       \item
        {\bf Traditional uploading}: The participants directly select and upload the file with their responses. 
        \end{itemize}
\end{enumerate}
We store the uploaded questionnaires and additional metadata, such as the method used for downloading and uploading and the browser's user agent. Finally, the participants are provided a compensation code that they input on Prolific to get paid €3. We note that we compensated all the participants who meaningfully participated, even when they could not upload their responses for various reasons (See the paragraph below). Hence, the users are not constrained to finalize all the steps before compensation. Therefore, we expect that this setup does not introduce biases in the provided responses and the security sense of participants in the study. 
We published the study on 01/15/2025 with an initial target of 50 participants. All users completed the survey within five hours. 
\\
{\bf Data cleanup.} 
We excluded users who did not submit their responses, e.g., due to network issues or when the uploaded file was too large.
We also excluded users who submitted a wrong PDF instead of their responses and those who used a mobile or non-Chromium browser. The results presented below are for the  N=\participantsfinal\ participants with valid responses. 
\\
{\bf Results.}
Table~\ref{tab:userstudyresultsumary} summarizes the main findings on how the participants interacted with the study website. 
Approximately 90\% of the users downloaded the questionnaire as a ZIP file, while 60\% uploaded their responses using the FSA API. We note that drag-and-drop is the preferred method for the users who uploaded using the FSA API (70\%). Most importantly, 50\% of the participants downloaded the questionnaire as a ZIP file and uploaded their responses via the FSA API, following the exact attack steps from Section~\ref{sec:fsa_vs_browser_profiles}. We note that users had safer alternatives to follow than providing write consent using the FSA API, suggesting that they are unaware of the dangers posed by this operation.
We note that eight participants reported being familiar with the FSA API. However, when asked if they found anything suspicious about the study/actions to be followed, most users responded that they found nothing suspicious to say, reported no, or \textit{I didn't even think of security}. Notably, one user says \textit{I was just concerned that I downloaded a zip file instead of filling out the form online}. 
\\
{\bf Further analysis of the responses.}
Though we requested users with Chromium browsers to participate in the study --with Chrome (37), Edge (23), and Opera (19)--, the users also regularly use Firefox (27) and Safari (26). This shows the feasibility of our cross-browser profile compromise. 
Interestingly, 15 participants reported already being familiar with browser profiles. The participants think that the profiles contain the following information: cookies (34), browsing history (30), bookmarks (28), passwords (26), auto-fill information (20), or user preferences (17). Users also mention extensions source codes (10) or root CAs (5).  As for where it is stored, only seven participants responded right for the local file system, while 14 think it is stored remotely in the Cloud, and 17 both remotely and locally. The persisted web data are considered moderately sensitive (20), highly sensitive (13), or not sensitive (3). Finally, 27 participants reported that they routinely install software from the web, which is concerning and might indicate susceptibility to malware attacks, in line with recent reports from Google~\cite{infostealingmalware}.
This shows that users have a fair understanding of what browser profiles contain and how they are stored on the disk, but they fail to understand the risks associated with persisting them on the disk.

%% file: countermeasures.tex
\section{Countermeasures}
\label{sec:countermeasures}
Considering all the presented attacks and our interactions with the browser vendors, we feel uneasy about their disregard for the file system attacker.
At runtime, browsers implement and provide numerous layered security mechanisms that enable safe browsing, e.g., the SOP, CORS, CSP, HTTPS~\cite{SOP, CORS, CSP} etc. We argue that similar mechanisms must be implemented for code and data stored on the disk. 
We advocate for a defense strategy in which all the different stakeholders, namely the user, browser vendors, and operating systems, must each play their role: browser vendors cannot keep on claiming that protecting against the file system attacker is not their mandate when they introduce APIs that give access to the file system. Browsers rely on operating systems for their underlying functionality; thus, specific OS-level mechanisms may help better secure browsers. Finally, users should play their necessary part via education, but they should not be blamed or kept solely responsible for issues related to poor browser APIs design and implementation.

\subsection{Browser Vendors}
Browsers should assume the worst about the underlying OS: on the same system may reside malicious programs with the same level of privileges, which might have read, write, or even execute capabilities to the file system. This type of attack is supported by the rise in the prevalence of supply chain attacks~\cite{DuanBJAXISL19,Ladisa-etal-22-USENIX}.
Vendors should build mechanisms or guidelines to defend against or detect such attackers and to empower security-conscious users. 
\\
{\bf Separating data and code.}
One of the fundamental principles in software security is to separate data from code, with code responsible for executing tasks and data acting as passive information. This separation minimizes potential risks associated with code manipulation or unauthorized access to sensitive data. Our attacks show that the lines between data and code blur because compromising browser profiles allows both of these objectives. 
%
This paradigm shift is akin to allowing applications to tamper with critical OS configurations, which is generally prohibited in conventional security practices. 
This highlights the need for robust isolation and access controls within browsers to ensure that data and code do not inadvertently interfere, thereby maintaining the security and integrity of the user's browsing experience.
\\
{\bf Improving current schemes.}
A couple of security mechanisms that browser vendors currently implement to protect specific data, i.e., cookies and passwords (See Section~\ref{sec:attacks_cookies}, and~\ref{sec:attacks_saved_credentials_and_passwords}) in the profile folders could be enhanced with authentication~\cite{HMAC}. For instance, before encryption, the origin of websites can be added to passwords and cookies and stored on disk. After decryption, the browsers can extract the origin from the encrypted passwords and cookies and match it against the origin of the current website on which they are to be set. A mismatch indicates that an attacker has copied the data from another origin. As previously discussed, integrity checks on browser extensions have various flaws in all Chromium browsers. We propose replacing them with stronger alternatives like encrypting all the profile folders. 
\\
{\bf Generalizing encryption.}
We recommend that profile folders be only stored encrypted on the disk and that the decryption process either involves the user's participation or uses some OS-specific key material that hinders portability. Technically, providing such a feature to users is possible. For instance, Electron, a Chromium and Nodejs-based framework, has a safe storage feature that applications can directly use. This feature derives the encryption key via OS features, e.g., the macOS KeyChain. Browsers already use such features to store the key for encrypting passwords. 
In a recent blog post~\cite{App-bound-enc-Windows}, the Chrome team announced plans to deploy this feature to encrypt cookies and other sensitive data.
In another blog post\cite{DBSC-Chromium}, the Chromium team describes DBSC, a prototype solution that prevents cookie theft by binding cookies to the device via asymmetric encryption and ephemeral keys that are stored securely on the device.  However, as described in Section~\ref{sec:attacks_saved_credentials_and_passwords}, attackers can still exploit how the encrypted data is stored on the disk, inside browser profiles. 
Thus, we think vendors should follow a similar scheme to encrypt entire profile folders, not only sensitive data items, and then store them in a transparent storage system, as they do now.
In Section~\ref{sec:evaluation_encryption}, we assess the feasibility and overhead of an encryption scheme on top of browser profiles. 
Another solution is for browser profiles to only be decrypted after a remote attestation step that guarantees that the profile belongs to the correct machine or that the user explicitly requested migration to a different machine. 
Once decrypted and to avoid storing them in clear on the disk, the profiles could, be stored in memory, e.g., using the OPFS API~\cite{OPFS}.

\subsection{Operating Systems}
Security-conscious users must consider supplementing the browser's security controls with additional OS-level ones. For macOS users, there exists a valuable feature that allows users to restrict applications' access to specific folders, as detailed in Apple's support documentation~\cite{MacFS}. This capability lets users finely control which apps can access their sensitive files and data, bolstering security.
On Linux, kernel modules like AppArmor~\cite{AppArmor} can provide additional protection by defining and enforcing application access policies. 

\subsection{An Encryption on Top of Browser Profiles}
\label{sec:evaluation_encryption}
\label{sec:encryption_browser_profiles}
\begin{figure}
    \centering
    \includegraphics[width=0.75\linewidth]{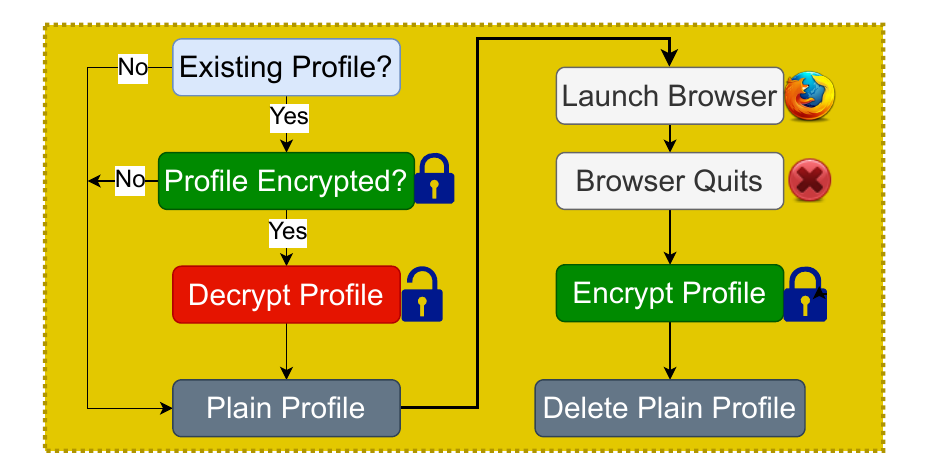}
    \caption{Encryption Scheme for Browser Profiles Security}
    \label{fig:encryption_browser_profiles}
\end{figure}

In this section, we report on the feasibility of encrypting entire browser profiles. Figure~\ref{fig:encryption_browser_profiles} summarizes the main steps of the proof of concept. The encryption logic is provided by GNUPG~\cite{GnuPG}.
When run, the prototype first checks whether a profile folder exists and whether it is encrypted. The user is then prompted to provide the key to decrypt the profile folder. The prototype then launches the browser with the plain profile. At the end of the regular browsing session, the user is prompted to provide the key to encrypt the disk's profile folder. This process is repeated when the user relaunches the program. The PoC includes detailed instructions for using the prototype with different browsers on Linux and macOS.
\\
{\bf Limitations.}
The main weakness of our prototype comes from the fact that, at various moments, the plain profile folder is visible on the disk, e.g., while the user is using the browser and before the encryption of the profile folder. A viable solution, which vendors might consider, would consist of keeping the plain profile folder in memory and only writing the encrypted profiles to the disk. In terms of performance, unsurprisingly, the encryption is more expensive than the decryption. In production, many aspects of the solution, particularly the performance, could be improved, but we argue that this is a viable direction that browser vendors could follow. However, repeated password requests might negatively influence the browser's usability. Thus, we leave for future work to explore the associated tradeoff further.

%% file: disclosure.tex
\section{Ethics, Disclosure and Open Science}
\label{sec:responsible_disclosure}
\noindent
{\bf Ethics.} 
The attacks were mounted in a controlled setting: only the authors' browsers and machines were involved in the experiments. The proofs-of-concept and other materials developed in this work have not been made public. The vendors to whom we reported the issues kept them private until they decided they had considered fixing the issues, in which case the bugs were publicly visible. 
We submitted and discussed the user study with the ERB (Ethical Review Board) and the Empirical Research Support (ERS) teams at our institution to ensure we followed ethical and privacy-preserving principles. We did not collect any personal or private user information, install any malicious software on their devices, or attempt to compromise their browsers or devices in any way. 
\\
{\bf Responsible Disclosure.}
\label{sec:disclosures}
We responsibly disclosed extensive details about our findings, i.e., the missed sensitive (browsing profile) folders, including demos and instructions to reproduce the issues, to the Chromium security team in particular and other vendors, whenever relevant.
We also provide an artifact along with this submission, containing detailed instructions on reproducing the attacks presented in this work. Finally, we took the opportunity of the reporting to raise the question of how the vendor considers the file system attacker.
\\
{\bf Chromium Security Team and Related Browsers}
We filed the issues considered in our threat model: symbolic links, missed folders in the blocklist, and the question of custom browser profiles. The first one discusses how symbolic links and junctions can be abused to bypass the blocklist and access sensitive folders on the user's machine.
The Chromium security team recognized the issue as \emph{a serious vulnerability} and that \emph{it seems reasonable to include the paths \code{\$HOME/.mozilla, \$HOME/.pki} in the blocklist}. For the \code{\$HOME/snap} folder, the question was \emph{if we should block the whole directory, or known subdirectories (though it would be impossible to come up with an exhaustive list)}. Furthermore, in a thorough revision
\emph{the loophole of providing a custom Chrome user directory has been fixed}, ensuring that custom profile folders are blocklisted as well. Regarding symbolic links, the Chrome security team confirmed that it is a critical security issue, but mentioned that they were already aware of it, hence deferring the discussion to the prior bug. Nonetheless, the changes they envisioned had to do with resolving the effective path of folders to ensure that they do not point to blocklisted ones. We also proposed changing the candidate specification of the FSA API to account for this subtle aspect of path resolution.
The browser team reported that there have been previous reports of such an issue, though the bugs are kept private and we could not access the issues. Since the prior bugs are not public, we think we can also be legitimately credited for raising the issue again. The issue is mitigated by resolving the real path when selecting a folder with the FSA API. The security team also agreed that the blocklist as a security mechanism is not the most effective one as bypasses will likely continue to be discovered. As of now, they agreed to add to the blocklist the missed folders, Firefox browser profile folders, and folders created by tools like Snap. As for custom profile folders, the Chromium sources have been revised to account for them. 
As for the integrity checks on signed extensions, we reported to the Chromium team an enforcement delay we observed on major Chromium browsers like Google Chrome when they perform integrity checks on manipulated extensions. This delay is long enough for malicious code to be executed before detection. The team reproduced the bug, but we are not aware if they took any further remediation steps. We also exchanged multiple emails with the Vivaldi team. They claim that this is an experimental feature, and contacted the Chromium Team to understand why it is not enabled by default in the sources and would abide by future changes made by Chromium. As for the file system attacker, they consider that this is the OS mandate and not that of the browser. We also reported the issue to Opera, but the vendor considered the issue a low profile because the attacker has access to the file system.   
\\
{\bf Firefox Security}
We filed an issue to disclose the bug we have reported to Chromium to discuss how Firefox is affected by the misuse of the FSA API, in a multi-browser setup. The team was very receptive to the discussion on considering and defending against the file system attacker. They have tried many attempts in the past, like encryption, and think of generalizing those to other parts of the profile folders. Usability and performance were natural questions as to where to keep the cryptographic keys and the potential performance hit of encrypting an entire profile folder.

\subsection{Compliance to Open Science}
To foster reproducibility, all the offensive and defensive materials developed as part of this work are publicly released~\cite{PoC}.

%% file: discussions.tex
\section{Limitations}
\label{sec:discussion}
\label{sec:limitations}
\label{sec:safari_browsers}
\label{sec:user_studies}
\label{sec:mobile_browsers}

We do not claim that the attacks discussed in this work are exhaustive. The files we have considered represent a tiny subset of the staggering amount of content that makes up profile folders. Yet, we have shown that this data is sufficient to mount powerful attacks that give attackers access to sensitive user information and bypass browsers' security mechanisms. All in all, there are many more attacks that we have inconclusively attempted, many attacks that can be mounted differently, and additional social engineering and involvement from the user could open new possibilities for the attacker. Finding and demonstrating new attacks can be rewarding. Nonetheless, our choices were to focus on high-profile attacks, to bring up scientific discussion about the new capabilities enabled by APIs like the FSA, and to question the security practices of browser vendors concerning storing user data on the file system. One possible improvement to be considered by browser vendors is to store the profiles encrypted using a device-bound key, similar to safeStorage in Electron\footnote{\url{https://www.electronjs.org/}}. However, this would make browser profiles not portable, negatively impacting usability for certain users.

%
\noindent
{\bf Safari browser.}
We note that OS mechanisms can significantly restrict the file system attacker. For example, macOS has many built-in access control mechanisms restricting various applications' access to the file system. For instance, the \code{Libray/} folder from the user's home directory, where many applications data is stored, is not visible by default when performing file browsing. 
Nonetheless, the Safari browser is also affected by many of the attacks discussed so far with a couple of differences. Its profile folder is located under \code{Library/Safari/}. 
WebExtensions are also supported but require user action to enable them in specific profiles. We are not aware of any security-related command line flag for this browser. Custom root CA can be managed in a GUI or by a super user. 
This being said, browsing history, the most visited websites, bookmarks, and downloads are stored clearly and can be readily exfiltrated. The attacker can be granted permission to access devices like the camera or microphone and the ability to send web push notifications. Website client-side data is also present in the folder. While form data and information related to passwords are also stored in the profile, we could not decode this information conclusively. They are encrypted and require user passwords to be read. 
\\
{\bf Mobile browsers.}
At the time of writing, no mobile browser supports the FSA API. Moreover, by design, mobile applications enjoy fewer privileges and flexibility regarding their ability to manipulate the file system freely compared to their desktop counterparts. Mobile platforms implement sandboxing and access control mechanisms where applications can only access the data they write, much like with the OPFS API~\cite{OPFS}.

%% file: relatedwork.tex
\section{Related Work}
\label{sec:related_work}


We are the first to systematically analyze the security implications of a file system attacker against browsers' profile folders.
We think this is a timely topic due to the introduction of APIs like FSA and the rise of supply chain attacks, which allow web attackers to deploy file system payloads. 
Oz et al.~\cite{Oz-etal-23-USENIX} were the first to showcase misuses of this API, but they claim that FSA API cannot be used to access browser profile folders. Conversely, we showcase end-to-end profile compromise attacks mounted by web attackers.
%
%
The only other work that glances at browser's persistence is that of Sanchez et al.~\cite{Sanchez-etal-20-CANS}, who propose an attack that reverse-engineers and bypasses the integrity verification implemented by Chromium-browsers to protect the browser's preferences. 
%
However, this work is limited to Chromium browsers and Chromium's integrity protection mechanism. We show that other browsers are also affected and that even web attackers can carry out such attacks. We also discuss many more attacks, such as session hijacking or MiTM.
%
\\
{\bf Browser Security.}
Traditionally, drive-by-download attacks were the main way of compromising web browsers by escaping the confinement of the sandbox. Cova et al.~\cite{Cova-etal-10-WWW} analyze malicious JavaScript using traditional machine learning models with hand-crafted features to detect such code.
Rieck et al.~\cite{RieckKD10} propose a more efficient detection strategy that uses static and dynamic JavaScript code features. More recently, Fass et al.~\cite{Fass0S19} show that camouflaging malicious code into benign support code can evade most such classifiers. Hardy et al.~\cite{HardyCKSSWGD14} note that politically motivated, targeted malware campaigns are prevalent. In response, vendors significantly increased their security efforts by frequently vetting their sandboxes and by deploying additional security mechanisms like CSP, CORS, or permissions. Researchers further discuss shortcomings of these mechanisms~\cite{LuoLHN19, Nomoto0SAM23, SnyderKELH23} and ways of testing their effectiveness~\cite{Hothersall-Thomas15,BernardoVVCSAM24,ShouKSB21,ZhouZWGLLP022,BrownSE20}. Moreover, they show that technologies like WebAssembly~\cite{0002KP20} or Service Workers~\cite{SquarcinaCM21} often introduce their own security problems.
Unlike prior work, we study a novel way of exploiting browsers by compromising their persistence layer.
\\
{\bf Malicious browser extensions.} Man-in-the-browser attacks can also be carried out by inadvertently installing malicious browser extensions. A plethora of recent work aims to detect such extensions~\cite{ErikssonPS22,KapravelosGCKVP14,PantelaiosNK20} or to protect the websites' integrity in the presence of malicious extensions~\cite{LouwLV08}. While such extensions also compromise the browser's integrity, they still require the user's consent to install. Kim et al.~\cite{KimL23} show that attackers can also exploit vulnerable extensions to achieve privilege escalation. 
Yu et al.~\cite{Yu0ZC23} present a sophisticated static analysis for finding problematic extensions. 
Agarwal~\cite{Shubham} detects browser extensions that inadvertently interfere with the security mechanisms of modern browsers.
In this work, we are the first to show how web attackers can stealthily install malicious extensions by compromising browser profiles. 
\\ 
{\bf Supply chain attacks.}
Malicious JavaScript code is also common in the context of supply chain attacks. Ladisa et al.~\cite{LadisaPMB23} propose a taxonomy of such attacks and show that attackers can inject malicious code in multiple stages of the open-source development process. Zimmermann et al.~\cite{zimmermann2019small} show that the effect of malicious code quickly propagates in the NPM ecosystem, affecting multiple transitive dependencies and their clients. Duan et al.~\cite{DuanBJAXISL19} advocate for a sophisticated detection technique integrating metadata, static, and dynamic analysis to flag malicious JavaScript code in open-source packages. We are unaware of prior work that studies how supply chain attacks can compromise browser profiles.



%% file: conclusion.tex
\section{Conclusion}
\label{sec:conclusion}
In this work, we present a series of end-to-end web attacks that compromise the confidentiality and integrity of stored browser profiles, including the use of JavaScript code running inside a browser. 
We show that these attacks can lead to serious consequences, such as session hijacking, password theft, or the installation of malicious browser extensions. 
Modern browsers mix sensitive user data with privileged extension code in unencrypted, easy-to-modify browser profiles. We recommend that browsers separate these concerns and store extensions' code in a separate, read-only folder. Moreover, we advocate for additional security controls that prevent reading and writing browser profiles from outside the browser's internal code.
%
Overall, we argue that the file system attacker is relevant for web security and should be considered more often by browser vendors, users, and future work. We warn that the File System Access API is a dangerous bridge that can allow web attackers to elevate their privileges and become file system attackers. Considering the catastrophic impact of unrestricted file system access, the existing security controls are probably insufficient (incomplete blocklist, flawed support for symbolic links). We show that this attack, coupled with the browser's careless storage of highly sensitive user data and extension code, leads to powerful man-in-the-browser attacks or identity theft. 
We argue that web security work should more often assume the possibility of a compromised browser. A example in this direction is Fidelius~\cite{EskandarianCBBF19}, which proposes using trusted hardware indicators and enclaves to ensure the confidentiality and integrity of user data, assuming a compromised browser.

%% file: appendix.tex
\section{Browsers Profile folders on Other Operating Systems}
\begin{table*}[!htp]
    \centering
    \begin{tabular}{|l|p{3.8cm}|l|l|l|l|}
        \hline
         \bfseries{Browser} & \multicolumn{3}{c|}{\bf Profiles Folder} \\ \hline
         & \bfseries{Linux} & \bfseries{Windows} & \bfseries{Mac OS} \\ \hline
        Google Chrome & \code{\tiny .config/google-chrome/} &  \code{\tiny AppData\symbol{92}Local\symbol{92}Google\symbol{92}Chrome\symbol{92}User Data\symbol{92}} & \code{\tiny Libray/Application Support/Google/Chrome/} \\ \hline
        Microsoft Edge & \code{\tiny .config/microsoft-edge/} &  \code{\tiny AppData\symbol{92}Local\symbol{92}Microsoft\symbol{92}Edge\symbol{92}User Data\symbol{92}} &  \code{\tiny Libray/Application Support/Microsoft Edge/} \\ \hline
        Vivaldi & \code{\tiny .config/vivaldi/} &  \code{\tiny AppData\symbol{92}Local\symbol{92}Vivaldi\symbol{92}User Data\symbol{92}} & \code{\tiny Libray/Application Support/Vivaldi/} \\ \hline 
        Opera & \code{\tiny .config/opera/} &  \code{\tiny AppData\symbol{92}Local\symbol{92}Opera Stable\symbol{92}} & \code{\tiny Libray/Application Support/com.operasoftware/} \\ \hline 
        Avast Secure & \code{\tiny N/A} & \code{\tiny AppData\symbol{92}Local\symbol{92}AVAST Software\symbol{92}Browser\symbol{92}User Data\symbol{92}} & \code{\tiny Libray/Application Support/AVAST Software/Browser/} \\ \hline 
        Yandex & \code{\tiny .config/yandex/} &  \code{\tiny AppData\symbol{92}Local\symbol{92}Yandex\symbol{92}YandexBrowser\symbol{92}User Data\symbol{92}} & \code{\tiny Libray/Application Support/Yandex/YandexBrowser/} \\ \hline 
        Brave & \code{\tiny .config/brave/} &  \code{\tiny AppData\symbol{92}Local\symbol{92}BraveSoftware\symbol{92}Brave-Browser\symbol{92}User Data\symbol{92}} & \code{\tiny Libray/Application Support/BraveSoftware/Brave-Browser/} \\ \hline 
        Chromium & \code{\tiny .config/chromium/}, \code{\tiny snap/chromium/common/chromium/} &  \code{\tiny AppData\symbol{92}Local\symbol{92}Chromium\symbol{92}User Data\symbol{92}} & \code{\tiny Libray/Application Support/Chromium/} \\ \hline \hline
        Firefox & \code{\tiny .mozilla/firefox/, snap/firefox/common/.mozilla/firefox/} &  \code{\tiny AppData\symbol{92}Roaming\symbol{92}Mozilla\symbol{92}Firefox\symbol{92}Profiles\symbol{92}} & \code{\tiny Libray/Application Support/Firefox/Profiles/} \\ \hline 
        LibreWolf & \code{\tiny .librewolf/} &  \code{\tiny AppData\symbol{92}Local\symbol{92}librewolf\symbol{92}Profiles\symbol{92}} & \code{\tiny Libray/Application Support/librewolf/Profiles/} \\ \hline 
        Waterfox & \code{\tiny .waterfox/}&  \code{\tiny AppData\symbol{92}Roaming\symbol{92}Waterfox\symbol{92}Profiles\symbol{92}} & \code{\tiny Libray/Application Support/Waterfox/Profiles/} \\ \hline         
        Safari &  &  & \code{\tiny Libray/Safari/} \\ \hline 
        
    \end{tabular}
    \caption{Common default browsers profile folders locations on Linux, Windows and Mac OSX.}
    \label{tab:study_desktop_browsers}
\end{table*}

\section{Browsers command line options}

\subsection{Gecko-based Browsers}
Firefox browsers command line options~\cite{FirefoxCommandLineOptions} can be found by navigating \code{about:config} in a Firefox browser. With that said, the security-related features are better customized using the web-ext NPM package~\cite{WebExtFirefox}, as shown in Listing~\ref{lst:firefox_launch_command_line_options_with_web_ext}. 
\begin{figure*}
\begin{lstlisting}[language=bash, caption={Launching Firefox browser with command line options that impacts security}, label={lst:firefox_launch_command_line_options_with_web_ext}]
web-ext run
    --network.proxy.http=127.0.0.1 # HTTP proxy server IP
    --network.proxy.http_port=8080 # HTTP proxy server port
    --network.proxy.ssl=127.0.0.1  # HTTPS proxy server IP
    --network.proxy.ssl_port=8080  # HTTPS proxy server port
    --browser.safebrowsing.downloads.enabled=false #
    --browser.safebrowsing.downloads.remote.block_potentially_unwanted=false
    --browser.safebrowsing.downloads.remote.block_uncommon=false
    --browser.safebrowsing.malware.enabled=false
    --browser.safebrowsing.phishing.enabled=false
    --source-dir=/path/to/extension # temporary extension to load with browser
    --firefox-profile=custom_profile # custom profile folder
    --target=/usr/bin/firefox # firefox binary, can switch to another browser 
\end{lstlisting}
\end{figure*}
web-ext is an official package from Mozilla for browser automation. Note also the indication of the user profile folder which the attacker can point to the user default browsing profile. This is optional and the browser will fall back to a temporary and empty profile.
Firefox browsers color in a shade of red the search bar when an unsigned extension is installed. Otherwise, the extension loads and executes in the background, with access to the user's current and ongoing sensitive browsing data. 
As the attacker can execute commands, it can also redirect all user traffic to a MiTM proxy where HTTPS communications can be viewed and manipulated. We have tried to leverage this ability to tamper with the verification process of extensions downloaded by the browser to permanently persist malicious codes in extensions regularly installed by users. We observed that all modifications of extension sources we attempted were successfully detected and reverted by the browsers. 

\subsection{Chromium-browsers command line options}

Listing~\ref{lst:chromium_launch_command_line_options} shows a bash command which launches a Chromium browser with common security-related command line flags~\cite{ChromiumCommandLineOptions, FirefoxCommandLineOptions}.
\begin{figure*}
\begin{lstlisting}[language=bash, caption={Excerpt of Chromium browsers command line flags~\cite{ChromiumCommandLineOptions}, and how they can be abused to bypass security. These flags are also leveraged by automation tools like Puppeteer~\cite{Puppeteer} or Playwright~\cite{Playwright} that build on Chrome Devtools Protocol~\cite{ChromeCDP}.}, label={lst:chromium_launch_command_line_options}]
/usr/bin/google-chrome 
    --user-data-dir=custom_profile #  (Custom) browsing profile folder 
    --disable-web-security # Disable the SOP: websites can access one another data using the fetch API for instance 
    --disable-extensions-except=/path/to/extension # Disable user-installed extensions
    --load-extension=/path/to/extension # Load an unsigned (malicious) extension
    --proxy-server=127.0.0.1:8080 # Proxy server IP and port where to direct HTTP requests r
    --ignore-certificate-errors-spki-list=VkGpwRbOwoi+7boIZMKGdL4289oJVRZCqH5ZuUR8V78= # base-64-encoding of the MiTM proxy root CA.  
\end{lstlisting}
\end{figure*}

\begin{figure*}
    \centering
    \includegraphics[width=0.99\linewidth]{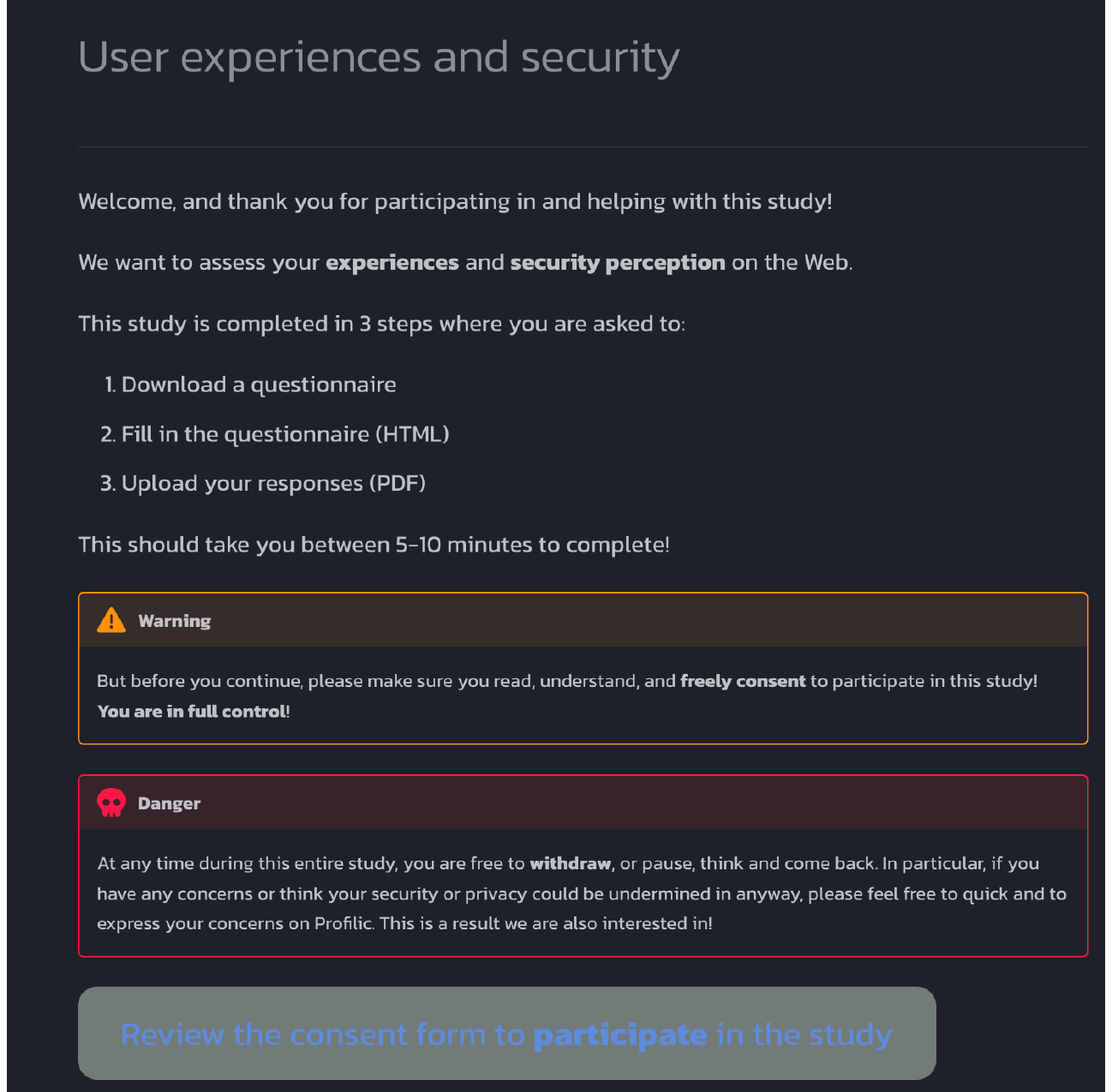} 
    \caption{User study website landing page}
    \label{fig:ustudy_landing_page}
\end{figure*}

\begin{figure*}
    \centering
    \includegraphics[width=0.99\linewidth]{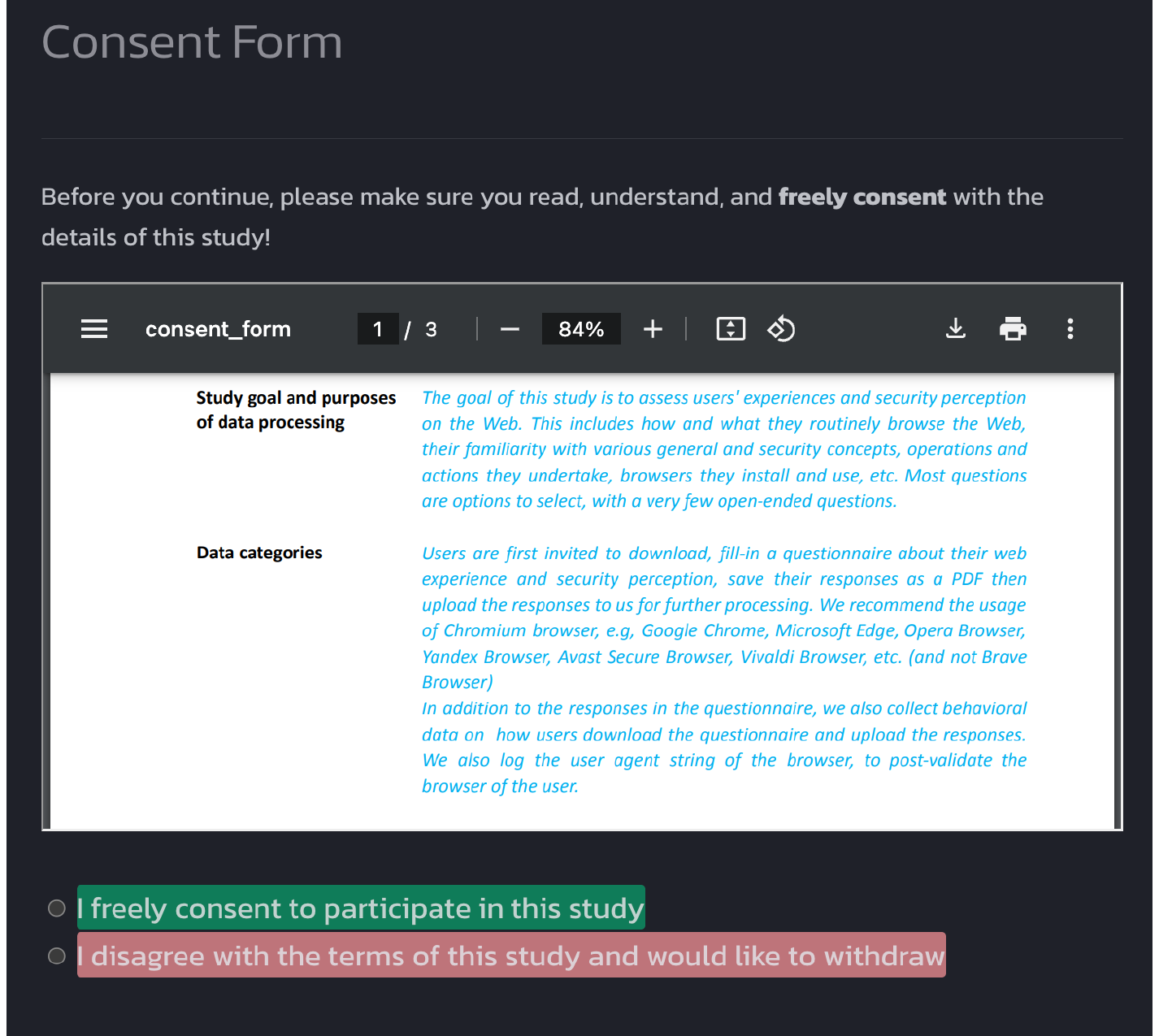} 
    \caption{User study consent form page}
    \label{fig:ustudy_consent_form}
\end{figure*}

\begin{figure*}
    \centering
    \includegraphics[width=0.99\linewidth]{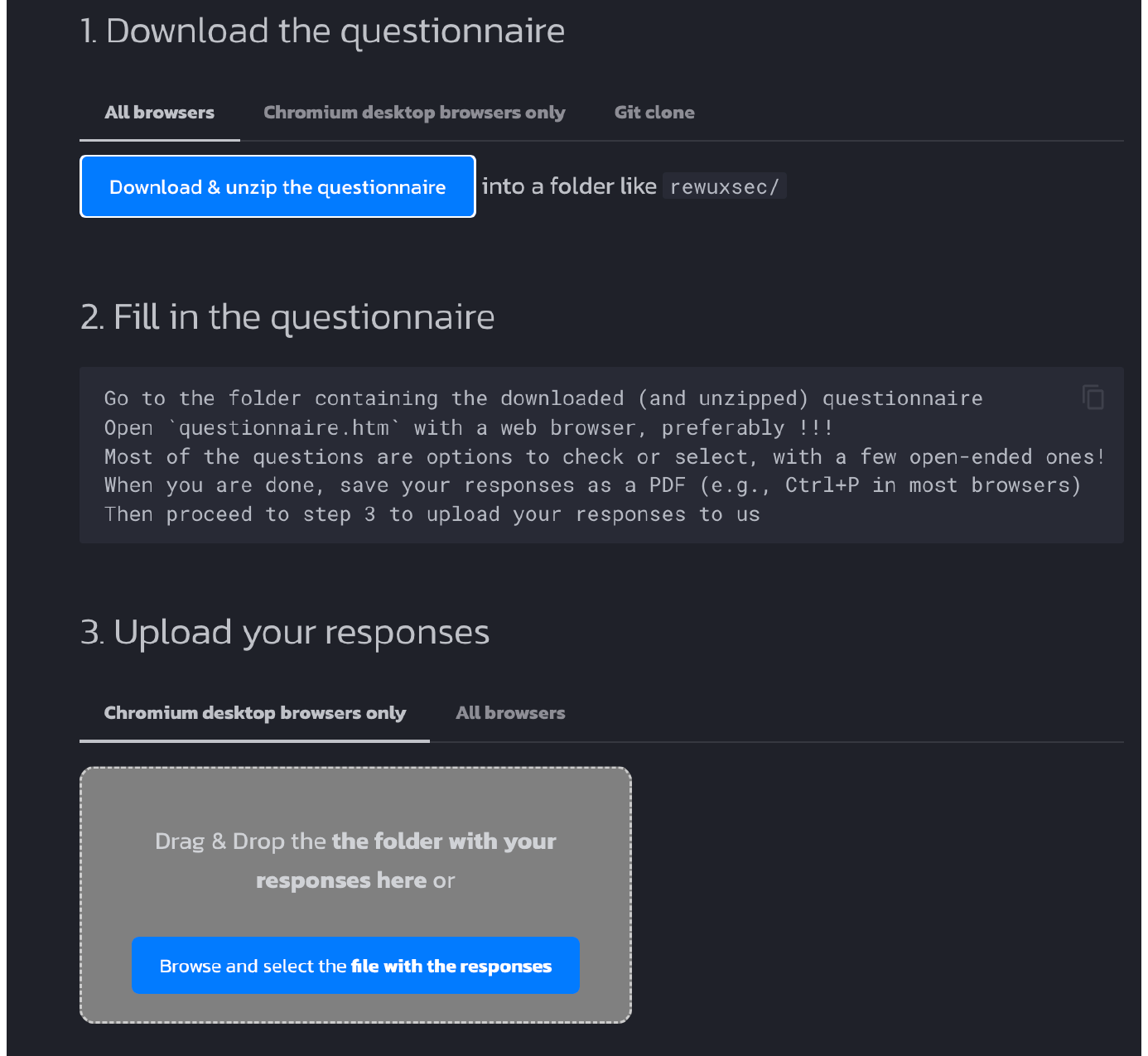} 
    \caption{User study steps to download the questionnaire, fill in and save it as a PDF, and upload the responses}
    \label{fig:ustudy_steps}
\end{figure*}


\newpage
\includepdf[pages=-]{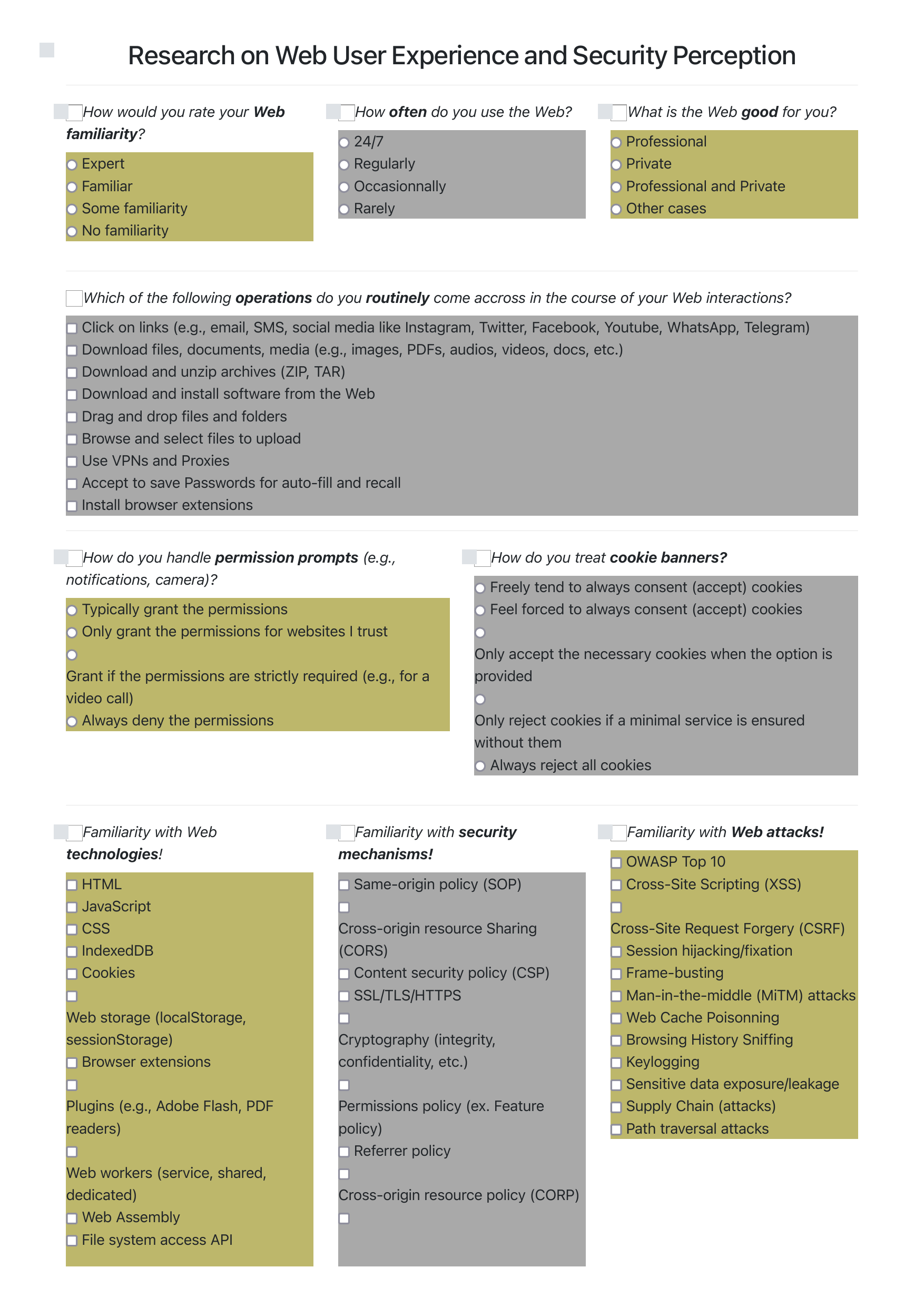}